\documentclass[12pt]{article}

\usepackage{amssymb}
\usepackage{amsmath}
\usepackage{graphics}
\usepackage{array}
\usepackage{afterpage}
\usepackage{epsfig}
\usepackage{float}
\newcommand{\te}[1]{{\text{#1}}}

\begin{document}

\title{\Large{Electron and Photon induced Reactions on Nuclei in the Nucleon 
Resonance Region\footnote{Work supported by DFG, GSI Darmstadt and BMBF}}}

\author{\large{J. Lehr, M. Effenberger and U. Mosel}\\
        \large{Institut f$\ddot{\textrm u}$r Theoretische Physik, 
        Universit$\ddot{\textrm a}$t Giessen}\\
        \large{Heinrich-Buff-Ring 16, D-35392 Giessen, Germany}\\
        \large{UGI-99-18}}
\date{}

\maketitle

\begin{abstract}
We calculate total photon-nucleus and electron-nucleus cross sections
$\gamma  A\to X$, $e A\to e^\prime X$ and several cross sections on pion and 
eta production in nuclei in the energy
regime between the first and the third resonance region for photon virtualities
$Q^2\le 1.0\ \textrm{GeV}^2$ within a semi-classical BUU
transport model. In both cases we discuss the varying influence of several
medium modifications on the cross sections for different values of $Q^2$.
\end{abstract}



\section{Introduction}

\noindent Photonuclear reactions are of interest for the investigation of the 
in-medium behaviour of hadrons. Especially the properties of the rho meson
in nuclear matter have been discussed in the last years, because they may be 
related to chiral symmetry \cite{koch}. 
Experiments on dilepton production in nucleus-nucleus collisions at SPS
energies \cite{agakichiev,mazzoni} seem to give evidence for a lowering of
the in-medium rho mass.
Measurements of the photoabsorption cross section on nuclei have shown a 
disappearance of the $D_{13}(1520)$ resonance \cite{bianchi}, which may be
explained by the changed rho properties \cite{post}. 

Furthermore, meson photoproduction reactions allow for a study of the 
in-medium dynamics of pions and etas, which
is closely linked to the behaviour of nuclear resonances at normal nuclear
matter density. However, due to the equality of energy and 
momentum of real photons, the 
different resonances can only be probed at fixed momentum transfer in the 
absorption process of the photon.

If one extends the discussion of photonuclear reactions to virtual photons,
one can get additional information on the in-medium dynamics of 
resonances, since energy and momentum can then be chosen independently from
each other. Electron-nucleus scattering yields a tool to study such 
processes, because the properties of the virtual photons can be determined by 
fixing the kinematics and the scattering angle of the electrons. Up to now, 
most electron scattering data are confined to reactions on nucleons, whereas 
measurements on nuclei only focussed on the process $eA\to e^\prime X$ (e.g. 
\cite{barreau,sealock}) for energies up to the $\Delta$ resonance. 

There are different discussions of inclusive and exclusive $(e,e^\prime)$
reactions on nuclei. In \cite{gil,gil2}, for example, there have been 
calculations within a $\Delta$-hole model, covering the energy regime of the 
quasielastic peak and the $\Delta$ resonance.

In the present work we consider electron-nucleus reactions from the 
$\Delta$ peak to the third resonance region. We extend the work presented in 
\cite{ef_abs,ef_prod} to virtual photons and also discuss some improvements 
of in-medium effects in photon-nucleus reactions.
Our calculations are based on a semiclassical BUU transport model 
\cite{ef_dil}, which has also been used very successfully in the past
for the description of heavy-ion collisions \cite{teis,hombach} up to energies 
of 2 AGeV and pion-nucleus reactions \cite{ef_pi}.

We proceed in section 2 by outlining the basic ideas of the BUU model.
In section 3 we show our parametrizations of the $\gamma^* N$ cross sections,
which we need to calculate the cross sections on the nucleus. 
In section 4 we present our results for the reactions $\gamma^* A\to X$ and 
discuss the influence of several medium modifications. We also compare our
calculations with data. Furthermore, we discuss modifications of the second
resonance region in $\gamma A$ and $\gamma^* A$ reactions.
In section 5 we turn to the photo- and electroproduction of pions and
etas on nuclei, showing total and differential cross sections on ${}^{40}$Ca.
We discuss the $Q^2$-dependence of the production ratios for pions and
etas on different nuclei. 
We close with a summary in section 6.

\section{The BUU Model}
 \label{sec:buu-model}
\noindent Most calculations presented in this work have been performed by
using a semi-classical BUU transport model. Here we just want to outline its
basic ingredients. For more detailed information the reader is referred to
\cite{ef_dil}.

The model is based upon the BUU equation
\begin{equation}
 \label{eq:buu}
\partial_tf(\vec r,\vec p,t)+{\partial H\over\partial\vec p}\vec\nabla_r f(\vec r,\vec p,t)-{\partial H\over \partial\vec r}\vec\nabla_p
f(\vec r,\vec p,t)=I_{coll}[f]
\end{equation}
with 
\begin{equation}
  \label{eq:mean-field}
  H=\sqrt{(m+S)^2+\vec p^2}.
\end{equation}
$f$ denotes the single-particle phase space 
density. For $I_{coll}\equiv0$ we obtain the Vlasov equation, which describes 
the time evolution of a many particle 
system under the influence of a mean field Hamilton function $H$. The 
right-hand side is the collision integral, consisting of a gain and a loss 
term, which accounts for possible collisions between the particles. 
For a system with different particle species one obtains a transport equation
for each species that is coupled to all others via the mean field and the 
collision integral.

Besides the nucleon the model includes in the non-strange sector the 
resonances 
$P_{33}(1232)$, $P_{11}(1440)$,
$D_{13}(1520)$, $S_{11}(1535)$, $P_{33}(1600)$, $S_{31}(1620)$,
$S_{11}(1650)$, $D_{15}(1675)$, $F_{15}(1680)$, $P_{13}(1879)$,
$S_{31}(1900)$, $F_{35}(1905)$, $P_{31}(1910)$, $D_{35}(1930)$,
$F_{37}(1950)$, $F_{17}(1990)$, $G_{17}(2190)$, $D_{35}(2350)$, 
which couple to the channels $N\pi$,$N\eta$,$N\omega$,$\Lambda K$,
$\Delta(1232)\pi$,$N\rho$,$N\sigma$,
$N(1440)\pi$ and $\Delta(1232)\rho$. 
Furthermore we explicitly propagate the mesons $\pi$, $\eta$, $\rho$, $\omega$,
$\phi$ and $\sigma$ (for the description of correlated $\pi\pi$ pairs with 
isospin zero). Meson-baryon collisions are mainly described through
excitations of intermediate baryonic resonances.
Above energies of $\sqrt s=2.6$ GeV (for baryon-baryon collisions) and 
$\sqrt s=2.2$ GeV (for meson-baryon collisions) we use the string fragmentation
model FRITIOF \cite{anderson}.

The initialization of the nucleon phase space density $f_N$ in
coordinate space is done according to a Woods-Saxon distribution 
$$
\varrho(r)=\varrho_0\left(1+\exp{r-r_0\over\alpha}\right)^{-1}
$$
with parameters $r_0$ and $\alpha$ that are fitted to experimental data.
In tab. \ref{tab:parameters} the parameters $r_0,\alpha$ are given for
the nuclei ${}^{12}\textrm{C}$, ${}^{40}\textrm{Ca}$ and ${}^{208}\textrm{Pb}$.

In momentum space we use local Thomas-Fermi approximation, which 
yields a local Fermi momentum
\begin{equation}
 \label{eq:th-fermi}
p_F(r)=\left({3\over 2}\pi^2\varrho(r)\right)^{1/3}.
\end{equation}
For the effective scalar potential $S$ appearing in eq.
(\ref{eq:mean-field}) we use the expression \cite{cassing}
\begin{equation}
 \label{eq:pot-mom}
S(\vec p,\varrho)=A{\varrho\over\varrho_0}+B\left({\varrho\over\varrho_0}
\right)^\tau+{2C\over\varrho_0}\sum_{I,S}\int{d^3p^\prime\over (2\pi)^3}
{f(\vec r,\vec p^\prime)\over 1+\left({\vec p-\vec p^\prime\over \Lambda}
\right)^2}
\end{equation}
with parameters $A,B,\tau$ fitted to binding energy, saturation density and 
compressibility \cite{teis}.
In this paper we use a hard (momentum independent) EOS with compressibility
$K=380$ MeV and $A=-124.3$ MeV, $B=71.0$ MeV, $C=0$, $\tau=2.0$ and a medium
momentum dependent EOS with $K=290$ MeV, $A=-29.3$ MeV, $B=57.2$ MeV,
$C=-63.5$ MeV, $\tau=1.76$ and $\Lambda=2.13\ \te{fm}^{-1}$.

To take into account that the $\Delta$ is less bound inside a nucleus than the
nucleons we use the $\Delta$ potential
\begin{equation}
  \label{eq:delta-pot}
  U_\Delta(p,\varrho)={2\over3}S(p,\varrho).
\end{equation}

The exact Thomas-Fermi groundstate corresponding to a local potential as in eq.
(\ref{eq:pot-mom}) is given by a step function in coordinate space:
$\varrho(r)=\varrho_0\Theta(R-r)$. Therefore an initialization of a realistic
density profile leads to oscillations. We have checked that our results are not
influenced by those oscillations by using a finite range Yukawa potential that
gave a practically stable nucleus. We use a local potential because for the 
calculation of relativistic heavy-ion collisions a numerical realization of a 
finite range potential is difficult.

\subsection{Treatment of $\Delta$ absorption}
 \label{sec:delta-in-med}

\begin{figure}[H]
  \begin{center}
\epsfig{file=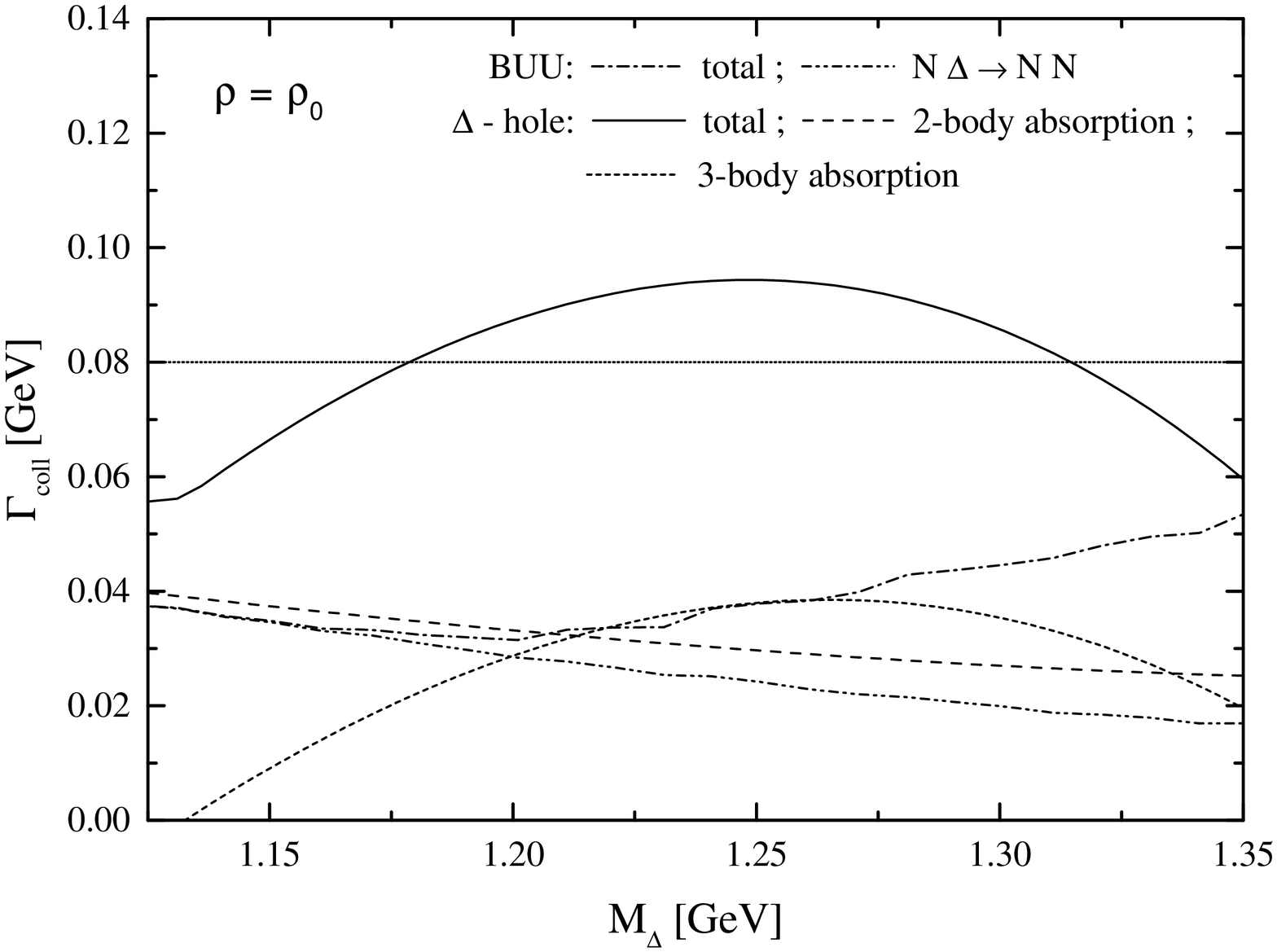,width=14cm}    
    \caption{Comparison of the different $\Delta$ in-medium widths. The curves 
labelled with 'BUU' correspond to the treatment in \cite{ef_abs}, the curves
labelled with '$\Delta$-hole' to that of \cite{oset_delta}. The quasielastic
contributions are not displayed, but included in the total widths.
Also shown is
the constant collision width extracted from the spreading potential.}
    \label{fig:coll_width}
  \end{center}
\end{figure}

\noindent In \cite{ef_abs} we have calculated the in-medium width of the
$\Delta(1232)$ resonance within our transport model. As usually done in
transport calculations, we have used vacuum cross sections and only took into
account two-body collisions. As a result the total in-medium width, consisting
of the spontaneous decay width and the collisional width, was very close to
the vacuum width, because the effects of the reduction of the $\pi N$ decay
width due to Pauli blocking and the collisional broadening nearly cancelled 
each other. However, these results are in contradiction to what one knows from
calculations within the $\Delta$-hole model \cite{hirata,weise}. There the
$\Delta$-'spreading potential' was from $\pi A$ scattering 
determined to be
\begin{equation}
  \label{eq:collision_width}
  -{\textrm {Im}}(V_{sp})={\Gamma_{coll}\over2}=(40\ \textrm{MeV}){\varrho\over\varrho_0}.
\end{equation}
The microscopic calculations of the $\Delta$-self energy in \cite{oset_delta} 
gave also results that are in line with this value.
In fig. \ref{fig:coll_width} we compare these collisional widths with our 
results. The widths are shown as a function of the $\Delta$ mass. One sees
that our total width (dash-dotted line) is about a factor two smaller than the
$\Delta$-hole value.
Compared to the $\Delta$-hole model calculation of \cite{oset_delta} it is in 
particular notable
that our result for the partial width from the process $N \Delta\to NN$
(dot-dot-dashed line) agrees rather well with what is called 'two-body' 
contribution (dashed line) in this calculation. The main difference to our
calculations therefore stems from the three-body contributions (short-dashed
line), which correspond to a process $\Delta NN\to NNN$ in a transport model.

In our calculations we did not explicitly include the three-body channel
but just determined a decay probability $p$ according to
$$
p=\exp(-\Gamma_{coll} t),
$$
where $\Gamma_{coll}$ here denotes the collisional width without the
quasielastic contribution. We have checked this procedure by
performing calculations in which we implemented the 'BUU' collision width
not via explicit collisions but through this method, which gave 
practically the same results.

As we will discuss below the momentum differential pion production cross
sections at photon energies above the $\Delta$ resonance are quite sensitive
to the treatment of the $\Delta$ absorption.

\section{Cross Sections for the $eN$ Process}
 \label{xsection}
\noindent Working in impulse approximation, we assume the electron to interact 
only with a single nucleon inside the nucleus via a virtual photon. The
differential cross section of the process $e N\to e^\prime X$ in the rest
frame of the nucleon is then given by:
\begin{equation}
\label{eq:elprod_xsec}
{d\sigma\over dE^\prime d\Omega}=\Gamma\sigma_{\gamma^*N\to X}=\Gamma(\sigma_T 
+\varepsilon \sigma_L).
\end{equation}
Here $\Gamma$ and $\varepsilon$ are the flux factor and the degree of 
longitudinal polarization of the virtual photons. Both quantities are functions
of the three independent variables $E,E^\prime$ (energies of the incoming and
outgoing photon) and $\vartheta$ (scattering angle):
\begin{align}
 \label{eq:gamma}
\Gamma &={\alpha\over 2\pi^2}{E^\prime\over E}{s-m_N^2\over 2m_N Q^2}
{1\over 1-\varepsilon}\\
 \label{eq:epsi}
\varepsilon &={1\over 1+2\tan^2{\vartheta\over 2}(1+{E-E^\prime\over Q^2})},
\quad\varepsilon\in[0,1].
\end{align}
$Q^2$ is the negative 4-momentum of the virtual photon:
\begin{equation}
 \label{eq:qsq}
Q^2=4E E^\prime\sin^2{\vartheta\over 2}.
\end{equation}
The cross section $\sigma_{\gamma^* N\to X}$ of the virtual photon interacting 
with the nucleon consists of contributions from all possible 
channels. Each one can be divided into two parts $\sigma_T$ and $\sigma_L$,
arising from transverse and longitudinal photons and can be 
written as functions of the photon variables $Q^2, E_\gamma=E-E^\prime$ and 
$\varepsilon$. In the limit $Q^2\to 0$, $\sigma_T$ equals the cross section 
$\sigma(\gamma N\to X)$ with a real photon in the initial state.
In our calculations we take into account one-pion, two-pion and eta 
production in elementary $\gamma^* N$ reactions. We use the cross sections
for $\gamma N$ described in \cite{ef_prod} to reproduce the energy dependence
and incorporate the $Q^2$ dependence by introducing form factors,
which are fitted to experimental data (see sec. \ref{sec:formf}).
 
Eq. (\ref{eq:elprod_xsec}) also contains the part $\sigma_L$, which
vanishes in the limit $Q^2\to 0$. It is known that the resonant contributions
to $\sigma_L$ are negligible \cite{stoler} and that $\sigma_L$ does not exceed
$0.2\cdot \sigma_T$ in the considered $Q^2$-range \cite{brasse}. Therefore,
the $\varepsilon$-dependence of the total $\gamma^* N$ cross section
(eq. (\ref{eq:elprod_xsec})) is very weak. 
Since there are yet no reliable values for $\sigma_L$ over a wide $Q^2$,
$\varepsilon$ and energy range available, we set
$\sigma_L=0$ and try to fit the form factors in such a way that $\sigma_T$
alone describes the total cross section $\sigma_{\gamma^* N\to X}$.
The weak $\varepsilon$-dependence is accounted for by choosing different fit 
parameters for the three bins $\varepsilon\ge0.9$, $0.6<\varepsilon<0.9$ and 
$\varepsilon\le0.6$.

\subsection{One-Pion Production}

$\sigma_T(\gamma^* N\to \pi N)$ can be written in terms of helicity amplitudes
\cite{arndt}:
$$
\sigma_T=\int d\Omega {\vert \vec p_\pi\vert\over 2 q_\gamma}\sum_{i=1}^4
\vert H_i\vert^2,
$$
where $\vec p_\pi$ is the pion momentum in the cms and $q_\gamma$ is
the equivalent photon momentum.
An expansion in terms of Legendre polynomials $P_l$ gives the following 
relationship between the $H_i$ and the transverse partial-wave amplitudes 
$A_{l\pm}$ and $B_{l\pm}$ ($j=l\pm1/2$, $l$: angular momentum of the 
$\pi N$-system):
\begin{equation}
 \label{eq:partialwellen}
 \begin{split}
  H_1 &={1\over\sqrt2}\sin\theta\cos{\theta\over2}
\sum_{l=1}^\infty(B_{l+}-B_{(l+1)-})(P_l^{\prime\prime}-
P_{l+1}^{\prime\prime})\\
  H_2 &={\sqrt2}\cos{\theta\over2}
\sum_{l=0}^\infty(A_{l+}-A_{(l+1)-})(P_l^{\prime}-P_{l+1}^\prime)\\ 
  H_3 &={1\over\sqrt2}\sin\theta\sin{\theta\over2}
\sum_{l=1}^\infty(B_{l+}+B_{(l+1)-})(P_l^{\prime\prime}+
P_{l+1}^{\prime\prime})\\
  H_4 &={\sqrt2}\sin{\theta\over2}
\sum_{l=0}^\infty(A_{l+}+A_{(l+1)-})(P_l^{\prime}+P_{l+1}^\prime).
\end{split}
\end{equation}
Here $\theta$ denotes the scattering angle in the cms.
The partial-wave amplitudes consist of a resonant and a background part.
This ensures that interferences between both contributions are taken into
account in the cross section.
Following \cite{ef_prod}, we use a Breit-Wigner ansatz for the resonant part:
 \begin{equation}
  \label{eq:ampl_ansatz}
  \begin{pmatrix}
A_{l\pm}(\sqrt s,Q^2)\\
B_{l\pm}(\sqrt s,Q^2)
\end{pmatrix}
=\begin{pmatrix}
A_{l\pm}(m_R,Q^2)\\
B_{l\pm}(m_R,Q^2)
\end{pmatrix}\cdot
{q^\te{R}_\gamma\vert\vec p_\pi^R\vert
\over q_\gamma \vert\vec p_\pi\vert}
{\Gamma_\te{tot}(m_R)\over
\Gamma_\pi(m_R)}\cdot {\sqrt s\ \Gamma_\pi^{1/2}
\Gamma_\gamma^{1/2}\over m_R^2-s-i\sqrt s\ \Gamma_\te{tot}}
\end{equation}
with
\begin{equation}
  \label{eq:ampl_res}
  \begin{split}
A_{l\pm}(m_R,Q^2) &=\mp F C_{\pi N}^I A_{1/2}(Q^2)\\
B_{l\pm}(m_R,Q^2) &=\pm F\sqrt{16\over(2j-1)(2j+3)}C_{\pi N}^I A_{3/2}
(Q^2)
\end{split}
\end{equation}
and
$$
F=\sqrt{{1\over\pi(2j+1)}{q_\gamma^R\over\vert\vec p_\pi^R\vert}
{m_N\over m_R}{\Gamma_\pi\over{\Gamma_\te{tot}^R}^2}}.
$$
$C_{\pi N}^I$ are the Clebsch-Gordan coefficients arising from the isospin
coupling of pion and nucleon. The superscript '$R$' denotes quantities
taken at the resonance mass. $\Gamma_\te{tot}$ and $\Gamma_\pi$ are the 
energy dependent total decay width and the one-pion decay width
of the resonance and are calculated 
according to 
\cite{manley}. $\Gamma_\gamma$ is a parametrization of the
$q_\gamma$-dependence of the $\gamma N$ width \cite{walker}.
Eq. (\ref{eq:ampl_ansatz}) leads to the resonance cross section
\begin{equation}
 \label{eq:res_xs}
\sigma_{\gamma^* N\to R\to\pi N}=\left({q_\gamma^R\over q_\gamma}
\right)^2
{s\Gamma_\gamma\Gamma_{R\to\pi N}\over (s-m_R^2)^2+s\Gamma_\te{tot}^2}
{2m_N\over m_R\Gamma_\te{tot}^R}(\vert A_{1/2}\vert^2+\vert A_{3/2}\vert^2).
\end{equation}

The photocoupling-helicity amplitudes $A_{1/2}$ and $A_{3/2}$ are functions
of $Q^2$, for which we make the ansatz 
\begin{equation}
\begin{split}
A_{1/2}(Q^2)=A_{1/2}(Q^2=0)\cdot f(Q^2)\\
A_{3/2}(Q^2)=A_{3/2}(Q^2=0)\cdot g(Q^2).
\end{split}
\end{equation}
Since the
amount of data points is rather limited, it is reasonable to keep the number 
of parameters as low as possible. Therefore we use the same form factor for
both amplitudes, although the $Q^2$-dependence of these quantities is known
to be different (e.g. \cite{warns}). Thus the form factors are fitted such that
they reproduce the $Q^2$-dependence of the expression 
$$
\vert A_{1/2}\vert^2+\vert A_{3/2}\vert^2,
$$
which equals the $Q^2$-dependence of the resonance cross section
in eq. (\ref{eq:res_xs}).

In our calculations the resonances $P_{33}(1232)$, $D_{13}(1520)$,
$S_{11}(1535)$ and $F_{15}(1680)$ are taken into account in elementary
$\gamma^* N$ interactions. For each of the partial waves, corresponding
to the four resonances, we introduce a different form factor. 
 
The background contributions to the partial-wave amplitudes were obtained
in \cite{ef_prod} by subtracting the resonance contributions 
(\ref{eq:ampl_ansatz}) from the partial-wave amplitudes in \cite{arndt}. To 
describe the $Q^2$-dependence, we introduce an additional fitting function
$f^{1\pi}(Q^2)$. Then the cross section reads:
\begin{equation}
\begin{split}
  \sigma_T &=\int d\Omega{\vert\vec p_\pi\vert\over 2 q_\gamma}
\biggl(\sum_{i=1}^4\biggl\vert\sum_{l=0}^3H_i^{l,\te{res}}(\vartheta,\sqrt s,
Q^2=0)\cdot
f^l(Q^2)+\\
&\qquad +H_i^\te{bg}(\vartheta,\sqrt s,Q^2=0)\cdot f^{1\pi}(Q^2)
\biggr\vert^2\biggr).
\end{split}
\end{equation}
Here $H_i^{l,\te{res}}$ stands for the resonant helicity amplitudes of partial 
wave $l$, i.e. for the contribution of the resonance with angular momentum $l$.

\subsection{Eta Production}

\noindent The cross section for eta production is parameterized under the 
assumption, that each elementary eta meson arises from the decay of an 
intermediate
$S_{11}(1535)$. The cross section is then given according to eq.
(\ref{eq:res_xs}):
\begin{equation}
 \label{eq:eta}
 \begin{split}
\sigma_{\gamma^* N\to\eta N}=\left({q_\gamma^R\over q_\gamma}
\right)^2 &
{s\Gamma_\gamma\Gamma_{S_{11}(1535)\to\eta N}\over (s-m_{1535}^2)^2+s\Gamma_\te{tot}^2} 
{2m_N\over m_{1535}\Gamma_\te{tot}^R}\times\\
&\times\vert A_{1/2}(Q^2=0)\cdot 
f^{S_{11}(1535)}(Q^2)\vert^2,
\end{split}
\end{equation}
where $f^{S_{11}(1535)}$ equals the form factor $f^l$ in the last section for $l=0$.

\subsection{Two-Pion Production}

\noindent In \cite{ef_prod} we have written the two-pion cross section 
$\gamma N\to\pi\pi N$ as an incoherent sum of a resonant and a background part.
The resonant part is given in analogy to eq. (\ref{eq:res_xs}). The 
difference between these contributions and the experimental data for $Q^2=0$ 
was treated as background. For finite $Q^2$, we introduce due to the absence
of data in all two-pion channels another fitting function $f^{2\pi}$ and use 
the following expression for the background:
\begin{equation}
 \sigma_{2\pi}^\te{bg}(Q^2)=\sigma_{2\pi}^\te{exp}(Q^2=0) \cdot 
f^{2\pi}(Q^2)^2 -\sum_R\sigma_{2\pi}^R
(Q^2=0)\cdot f^R(Q^2)^2.
\end{equation}
The parameters of $f^{2\pi}$ can be determined by fitting to the total
cross section $\sigma(\gamma^* p\to X)$, since the two other main contributions
$\gamma^* N\to\pi N,\ \gamma^* N\to\eta N$ are already known (cf. next 
section).

\subsection{Formfactors}
 \label{sec:formf}
\noindent We now have to determine the four resonance form factors and the
two functions $f^{1\pi}, f^{2\pi}$. We restrict ourselves to the range
$Q^2\le1.0\ \textrm{GeV}^2$ and use the following ansatz:
$$
f(Q^2)=\left(1+\left(Q^2\over a\right)^b\right)^{-c}
$$
with parameters $a,b,c$.
Although the amount of experimental
data is rather limited in the considered energy and $Q^2$ range, it is
nevertheless possible to find a set of parameters with which most data can
be described in a satisfactory way. 
For the fitting procedure we used data from the exclusive channels 
$\gamma^* p\to \pi^0 p$, $\gamma^* p\to\pi^+ n$, $\gamma^* p\to p\eta$
and measurements of the helicity amplitudes $A_{1/2}, A_{3/2}$ for the
resonances. The parameters are displayed in tab. \ref{tab:ff_par}.
We stress here that we do not aim at a perfect extraction of resonance
form factors, but only at a good description of the $Q^2$ dependence of 
($\gamma$, meson) reactions on the nucleon.

\begin{figure}[H]
 \begin{center}
\epsfig{file=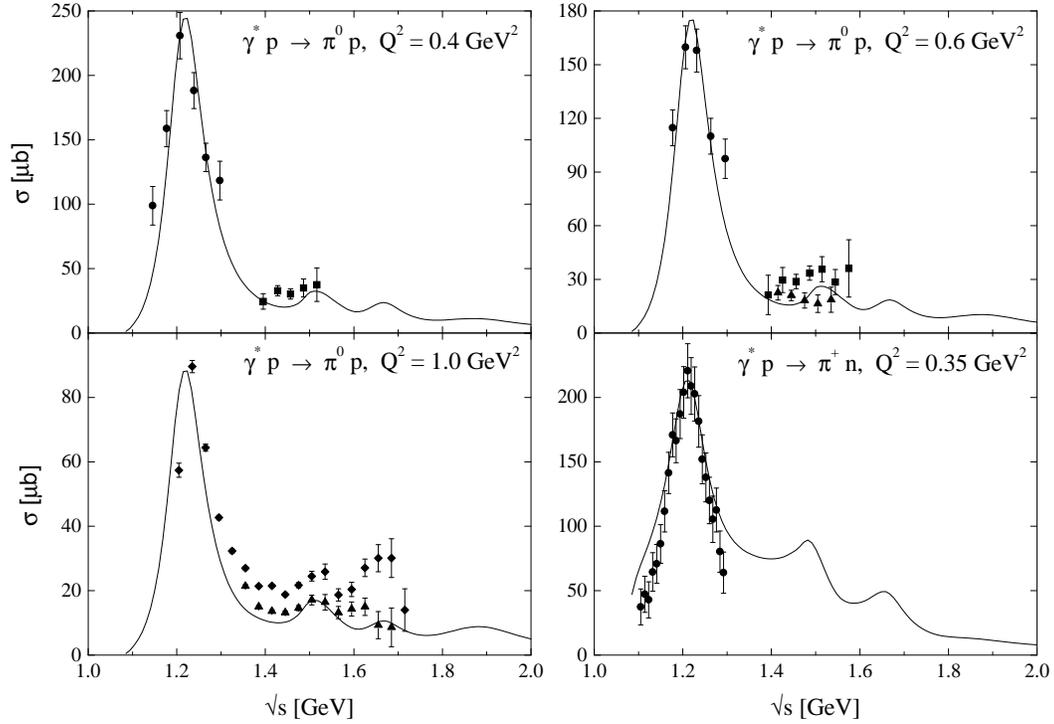,width=14cm}
     \caption{Comparison between pion electroproduction data and our 
calculations for different values of $Q^2$. a) - c): Cross sections
of the reaction $\gamma^* p\to \pi^0 p$ reaction for different $Q^2$. The data 
are from \cite{sh72} (circles), \cite{si71} (squares), \cite{al76}
(triangles) and \cite{la81} (diamonds). d): Cross section of 
the reaction $\gamma^* p\to \pi^+ n$-cross section. The data are from 
\cite{ga72}.}
     \label{fig:pi0p-pipn}
\end{center}
\end{figure}

\begin{figure}[H]
  \begin{center}
\epsfig{file=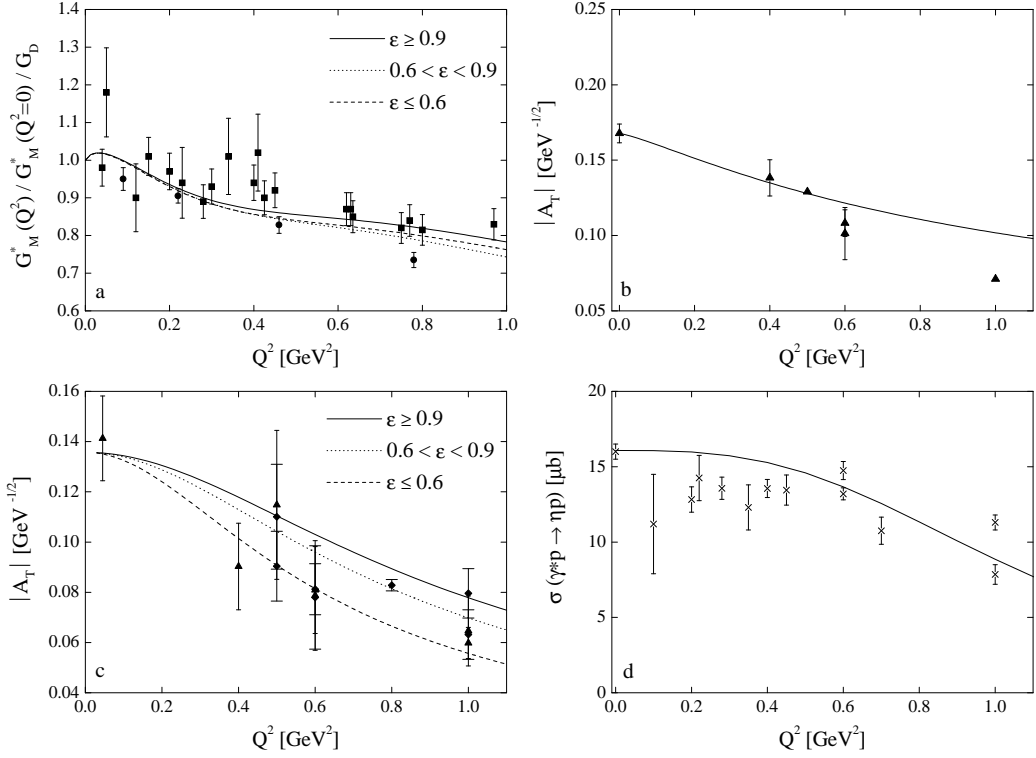,width=14cm}    
    \caption{a) Magnetic form factor of the $\Delta(1232)$. The  
curves correspond to different $\varepsilon$-bins, as indicated in the legend.
The data are from \cite{amaldi} (squares) and \cite{stein} (circles).
b) Transverse helicity amplitude $\vert A_T\vert$ for the $N(1520)$.
The data are from \cite{warns}.
c) Transverse helicity amplitude $\vert A_T\vert$ for the $N(1680)$. The curves
correspond to the $\varepsilon$-bins shown in the legend. 
The data are from \cite{warns} (triangles) and \cite{stoler} 
(diamonds).
d) Cross section of the reaction
$\gamma^* p\to\eta p$ for $\sqrt s=1.535$ GeV.
The data are from \cite{brasse78}. The data point for $Q^2=0$ is
from \cite{krusche}} 
    \label{fig:resonanz_ff}
  \end{center}
\end{figure}

We show the comparison of our calculations
with data from the channels $\gamma^* p\to\pi^0 p$ for $Q^2=0.4,0.6$ and
$1.0\ \te{GeV}^2$ and $\gamma^* p\to \pi^+ n$ for $Q^2=0.35\ \te{GeV}^2$ in
fig. \ref{fig:pi0p-pipn}.
In fig. \ref{fig:resonanz_ff} we compare with the $Q^2$-dependence of the
transverse helicity amplitudes $\vert A_T\vert=(\vert A_{1/2}\vert^2+\vert A_{3/2}\vert^2)^{1/2}$ of the resonances $D_{13}(1520)$ and $F_{15}(1680)$. For the
$\Delta(1232)$, we show in fig. \ref{fig:resonanz_ff} a) the magnetic form 
factor \cite{stoler}
$$
\vert G^*_M(Q^2)\vert^2=
\left({(Q^2+m_\Delta^2-m_N^2)^2\over 4m_N^2}+Q^2\right)^{-1}{m_N\over 2\pi\alpha}(m_\Delta^2-m_N^2)
(\vert A_{1/2}\vert^2+\vert A_{3/2}\vert^2).
$$
All these quatities do not depend on $\varepsilon$.
The three different curves for the $\Delta(1232)$ and the $F_{15}(1680)$ in
fig. \ref{fig:resonanz_ff}
arise from 
the necessity of using different fit parameters for the three 
$\varepsilon$ bins mentioned earlier. It can be seen that all three curves
are consistent with the data.
The cross section $\gamma^* p\to \eta p$ is dominated by the $S_{11}(1535)$ 
resonance and therefore reflects the $Q^2$-dependence of the helicity 
amplitude of this resonance (cf. eq. (\ref{eq:eta})).
 
\begin{figure}[H]
  \begin{center}
\epsfig{file=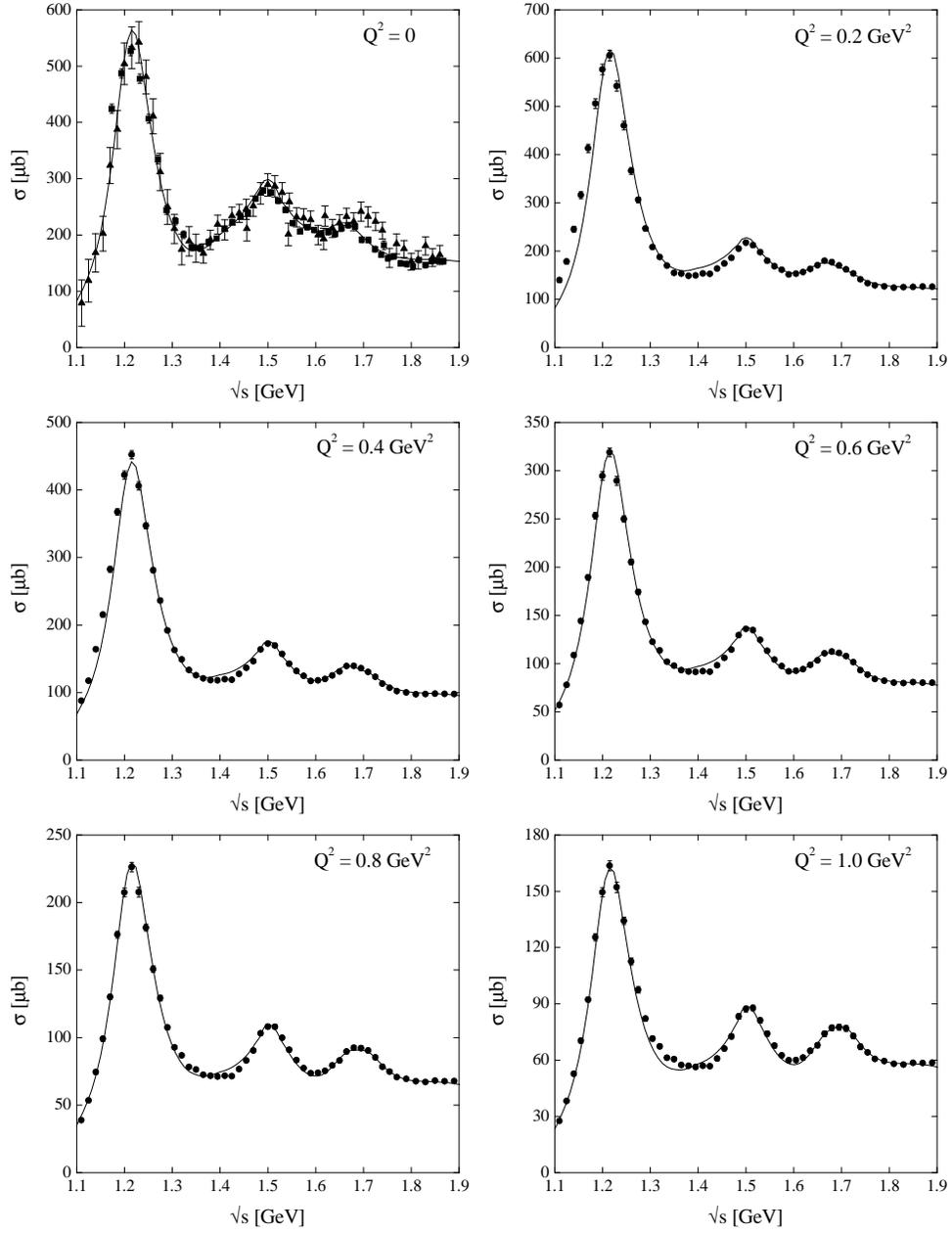,height=17cm}
    \caption{Total cross section of the reaction $\gamma^* p\to X$ for 
$\varepsilon\ge0.9$
and different $Q^2$. For $Q^2=0$ the data are from \cite{armstrong} 
(squares) and \cite{baldini} (triangles). For finite $Q^2$ the 
data are from \cite{brasse}.}
    \label{fig:xsection}
  \end{center}
\end{figure}

The remaining functions $f^{1\pi}$ and $f^{2\pi}$  were determined
by fitting the parameters to the total $\gamma^* p\to X$ cross section in 
\cite{brasse} in the first and the second
resonance region, respectively.
This leads to an underestimation of the cross section 
in the third resonance region of up to $25\mu$b. This difference 
has also been parametrized and treated as two-pion background.
In fig. \ref{fig:xsection} we show the total cross section $\gamma^* p\to X$
for different values of $Q^2$ and $\varepsilon\ge0.9$. For the two other
$\varepsilon$ bins, we obtain similar good agreement.

Since there are no data available for the different channels of the 
$\gamma^*n\to X$ cross section, we use the same form factors
for both protons and neutrons.

\section{The Processes $e A\to e^\prime X$ and $\gamma A\to X$}

\noindent In this section we are concerned with the process 
$e A\to e^\prime X$ and the photoabsorption process $\gamma A\to X$, which has
already been investigated in \cite{ef_abs}. Using the 
elementary medium-modified cross sections on the nucleons derived in the last 
section, the differential cross section of this process can be written in the 
form:
\begin{equation}
  \label{eq:el_kern_abs}
{d\sigma_{eA\to e^\prime X}\over d{E_e^\prime}d\Omega}
= 4 \int d^3r\int^{p_F}{d^3 p_N\over (2\pi)^3}\left(
{Z\over A}{d\sigma_{ep\to e^\prime X}\over d{E_e^\prime}
d\Omega}
+{N\over A}{d\sigma_{en\to e^\prime X}\over d{E_e^\prime}
d\Omega}\right),
\end{equation}
with
$$
{d\sigma_{e(p,n)\to e^\prime X}\over d\Omega_e dE_e^\prime}=
{d\sigma_{e(p,n)\to e^\prime\pi N}\over d\Omega_e dE_e^\prime}+
{d\sigma_{e(p,n)\to e^\prime\pi\pi N}\over d\Omega_e dE_e^\prime}+
{d\sigma_{e(p,n)\to e^\prime\eta N}\over d\Omega_e dE_e^\prime}.
$$
The Fermi momentum $p_F$ is given by eq. (\ref{eq:th-fermi}).
Note that the cross section has to be calculated in the rest frame of the
nucleus, so we have to Lorentz-transform the cross sections on the nucleons 
appearing in (\ref{eq:el_kern_abs}), since they were derived in the rest frame
of the nucleons. This can easily be done by using Lorentz invariance of the
expression $d^3p/E$ for the outgoing electron. Since the general expression
for the differential cross section $d\sigma={\vert{\cal M}\vert^2}
d\Phi/j$ ($j$: flux of incoming particles) involves the Lorentz invariant 
(matrix element)$^2$$\times$ phase space,
the connection between the nucleon cross section in the rest frames of the
nucleus (LAB) and the nucleon (R) is given by:
$$
\left({d\sigma\over dE^\prime d\Omega}\right)_{LAB}=
{j^R\over j^{LAB}}{{E^\prime}^{LAB}\over {E^\prime}^R}\left({d\sigma\over dE^
\prime d\Omega}\right)_R.
$$
In order to compare the cross sections of the processes $eA\to e^\prime X$ and
$\gamma A\to X$, one can define a cross section of the process
$\gamma^* A\to X$ according to eq. (\ref{eq:elprod_xsec}) by
$$
\sigma_{\gamma^* A\to X}={1\over\Gamma}{d\sigma_{e A\to e^\prime X}\over
dE^\prime d\Omega}
$$
with the flux factor $\Gamma$ from eq. (\ref{eq:gamma})
\cite{barreau}.

\begin{figure}[H]
  \begin{center}
\epsfig{file=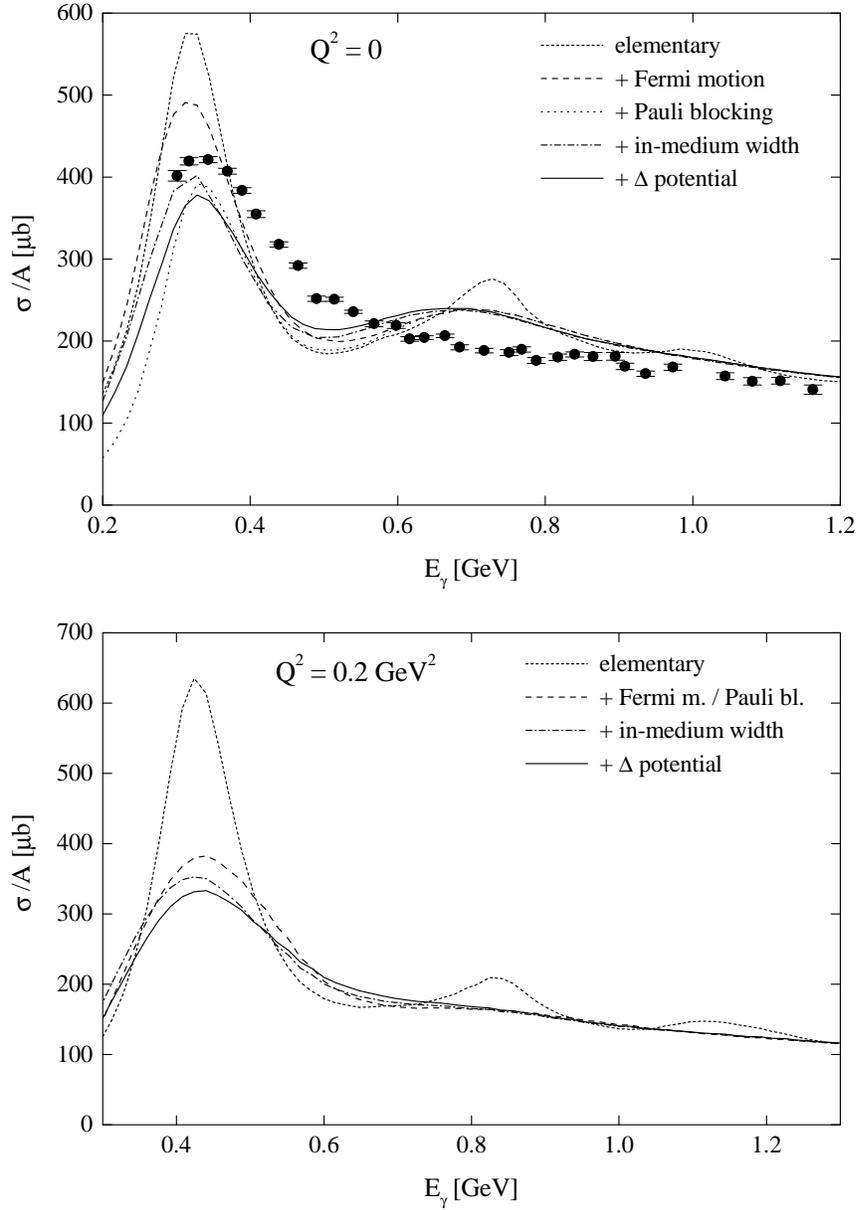,height=17cm}    
    \caption{Influence of different in-medium modifications on the cross 
section of the reaction $\gamma^* {}^{40}\textrm{Ca}\to X$ for different
$Q^2$ and $\varepsilon\ge0.9$. The different modifications are switched on
subsequently. All contributions due to quasieleastic scattering have not
been taken into account. The data are from \cite{bianchi} and were obtained
from an average over different nuclei.}
    \label{fig:absorption}
  \end{center}
\end{figure}

We now turn to the influence of the different in-medium modifications to the
cross section for different $Q^2$. In fig. \ref{fig:absorption} the case 
$Q^2=0$, which was discussed in detail in \cite{ef_abs}, is shown. 
The different curves show the different modifications which are turned on 
subsequently, as indicated in the legend. The elementary cross section without 
any modifications shows the three resonance regions. After turning on the 
Fermi motion of the nucleons the 
three resonance regions become smeared out, which causes the 
disappearance of the $F_{15}(1680)$.
Pauli blocking leads to a slight decrease of the cross section in the
$\Delta$ resonance region and a shift of the resonance maximum to larger 
energies.
The inclusion of in-medium widths, i.e. Pauli-blocked one-pion widths and the
additional collision width given by eq. (\ref{eq:collision_width})
results in a shift of the $\Delta$ peak to smaller energies and a slight 
increase,
which can be explained by the fact that the ratio of in-medium and vacuum
width of the $\Delta$ for energies up to the resonance mass roughly equals 1, 
whereas it increases quickly for energies above. The increase is due to the
additional possible collision reactions of the $\Delta$ in the nucleus.

When we also consider that the $\Delta$ is less bound in the nucleus than the
nucleons by using the $\Delta$ potential from eq. (\ref{eq:delta-pot})
(involving the momentum independent nucleon potential),
the increasing $\Delta$ mass causes a shift of the first resonance region back
to higher energies. The cross section again drops, since the $\Delta$ width 
increases with the mass. 
 
After applying all medium modifications, one observes a discrepancy with the
data from \cite{bianchi}
in the $\Delta$ region which was already discussed in \cite{ef_abs} and
could be cured by the inclusion of a two-body absorption process 
$\gamma NN\to N\Delta$ \cite{carrasco}.

\medskip

At finite $Q^2$, there are additional scattering processes at energies below 
the $\Delta$ mass, which  lead to the quasielastic peak. We do not consider 
these
processes in our investigations. The quasielastic region is discussed in 
detail in \cite{gil}, where the authors calculate the $eA\to e^\prime X$ 
reaction within a $\Delta$-hole model in the energy range up to the $\Delta$
resonance. The results on ${}^{12}$C and ${}^{208}$Pb show a good global
agreement with the data.

\begin{figure}[H]
  \begin{center}
\epsfig{file=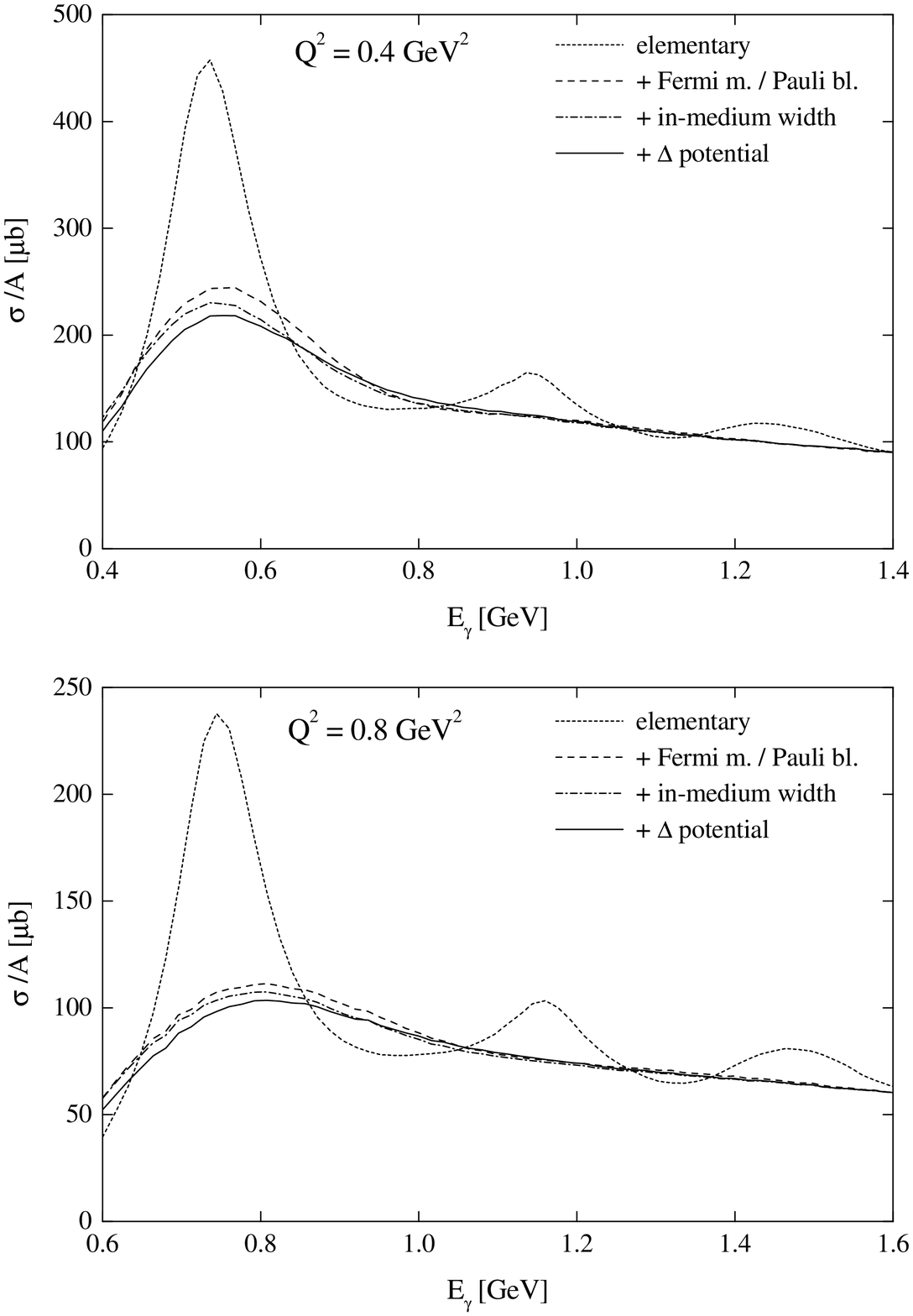,height=17cm}    
    \caption{Influence of different in medium modifications on the cross 
section of the reaction $\gamma^* {}^{40}\te{Ca} \to X$ for different $Q^2$
and $\varepsilon\ge0.9$.
The different modifications are switched on subsequently. All contributions 
due to quasielastic scattering have not been taken into account.}
    \label{fig:absorption2}
  \end{center}
\end{figure}

As can be seen in figs. \ref{fig:absorption} and \ref{fig:absorption2}, the 
influence of the different in-medium 
modifications changes with increasing $Q^2$. First of all, we observe that the
Fermi motion of the nucleons becomes more effective with increasing $Q^2$.
The cms energy is given by
\begin{equation}
  \label{eq:centerofmass}
s=-Q^2+m_N^2+2E_\gamma\sqrt{m_N^2+\vec p_N^2}-2\sqrt{Q^2+E_\gamma^2}p_N^z.
\end{equation}
The last term in this equation becomes more important for increasing $Q^2$,
which causes the mentioned effect and leads to a broadening of the resonant 
contributions and a disappearance of the second resonance region already for
$Q^2\ge 0.2$ GeV$^2$. Pauli blocking is totally suppressed for
$Q^2\ge 0.2$ GeV$^2$ (both curves lie upon each other), because for increasing
$Q^2$ the excitation of a certain resonance requires a larger momentum
transfer, which leads to a rising nucleon momentum after decay.
The consideration of in-medium widths results in a curve which lies 
below the one with Pauli blocking, different from the 
real photon case: The inclusion of the total in-medium width leads to a
dropping of the one-pion contributions to the cross section because of the
$1/\Gamma$ dependence of the resonant part (cf. eq. (\ref{eq:res_xs})).
For $Q^2=0$ this decrease is compensated by the additional collision
processes. For finite $Q^2$, one has to take into account the different
$Q^2$ dependencies of both contributions. In the collision part it is
totally determined by the $\Delta$ form factor, whereas the one-pion 
contributions also involve the background processes, generating a smoother
decrease for larger $Q^2$. Therefore the collision contributions become 
less important for increasing $Q^2$.

The $\Delta$ potential leads to a decrease of the cross section
as before.
The main observation is that the Fermi motion becomes the most important 
(though trivial) in-medium modification at large $Q^2$.
\medskip

\begin{figure}[H]
  \begin{center}
\epsfig{file=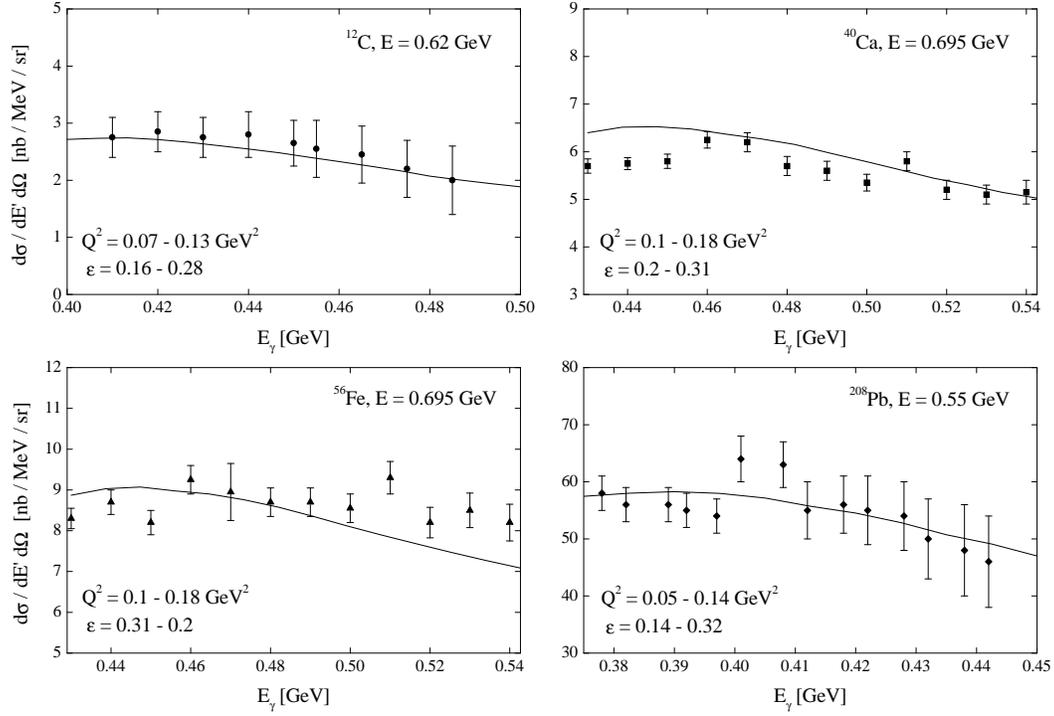,width=14cm}    
    \caption{Cross section of the reaction $eA\to e^\prime X$ measured at
$\vartheta=60^\circ$ on different nuclei for different initial electron
energies. The considered energy range corresponds to the upper $\Delta$
region. The $Q^2$ and $\varepsilon$ ranges are shown in the plots. The data
are from \cite{barreau} (circles), \cite{meziani} (squares and triangles)
and \cite{zgiche} (diamonds).}
    \label{fig:comp_data1}
  \end{center}
\end{figure}

We now compare our calculations with data 
measured in the energy range of the the upper $\Delta$ resonance region.
In fig. \ref{fig:comp_data1} we present our results for 
$\sigma(eA\to e^\prime X)$ for different nuclei. The electron energies 
$E\approx 0.6-0.7$ GeV and the scattering angle $\vartheta=60^\circ$  
correspond to rather small values of $Q^2$ in the range 
of $0.05-0.2\ \te{GeV}^2$. Since $E$ and $\vartheta$ are fixed, $Q^2$ and
$\varepsilon$ vary as $E_\gamma$ changes (see eqs. (\ref{eq:epsi}),
(\ref{eq:qsq})).
In each plot we show the ranges of $Q^2$ and $\varepsilon$.
We find excellent agreement with the experimental data in all cases.

\begin{figure}[H]
  \begin{center}
\epsfig{file=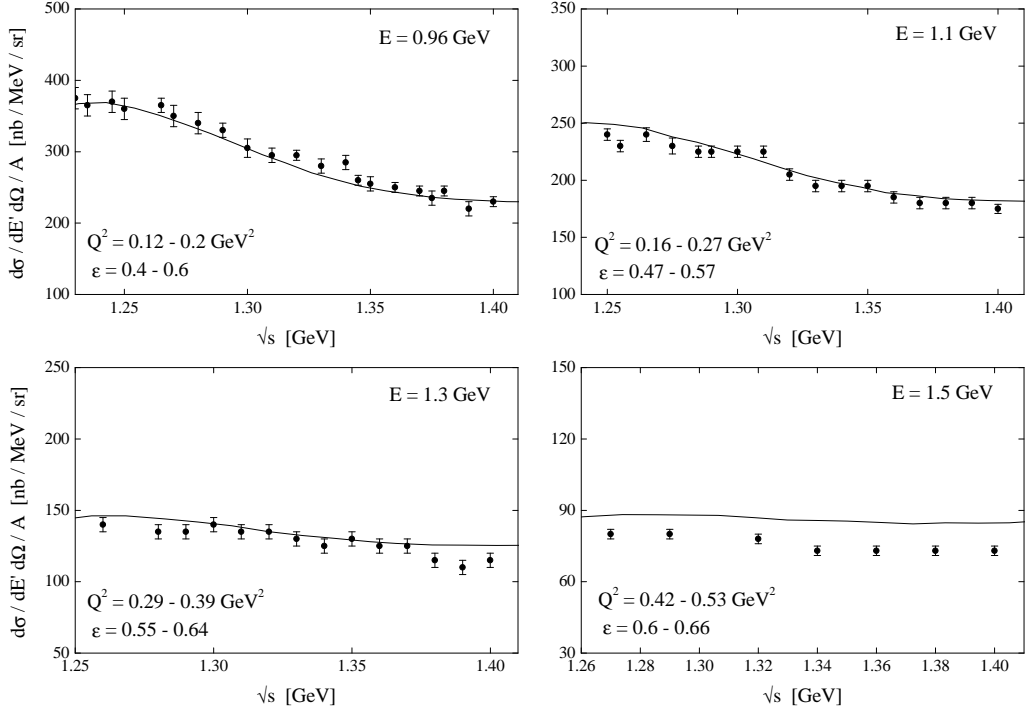,width=14cm}    
    \caption{Cross section of the reaction $e {}^{12}\textrm{C}\to e^\prime X$
measured at $\vartheta=37.5^\circ$ for different initial electron energies.
The energy range corresponds to the upper $\Delta$ region. The $Q^2$ and
$\varepsilon$ ranges are shown in the plots. The data are from \cite{sealock}.}
    \label{fig:comp_data2}
  \end{center}
\end{figure}

In fig. \ref{fig:comp_data2} we show results on ${}^{12}$C for different
values of $Q^2$. Again, the comparison with the data shows very good agreement,
although at $E=1.5$ GeV we overestimate the data by about 15\%. 
\medskip

So far we have seen that our model seems to be able to describe $eA$-reactions 
at least in the upper $\Delta$ resonance region. Additional measurements -
especially in the higher resonance regions - is needed to finally determine
whether this holds in the whole considered energy- and $Q^2$-region. 
The improved agreement in the $\Delta$ region compared to the case $Q^2=0$ 
without the inclusion of two-body absorption processes originates from the 
reduced de Broglie 
wavelength of the photon for finite $Q^2$. Therefore, the 
absorption process on single nucleons, as assumed in impulse approximation, is 
the dominating one.

\subsection{Medium Modifications of the $D_{13}(1520)$}

In fig. \ref{fig:absorption} one sees that for $Q^2=0$ 
the second resonance region is still visible, whereas the data 
clearly show a vanishing of any resonant structure. 
As discussed in \cite{ef_abs} the structure in this region is not only 
caused by the excitation of the $D_{13}(1520)$, but also by the opening of the 
two-pion channel. 

\begin{figure}[H]
  \begin{center}
\epsfig{file=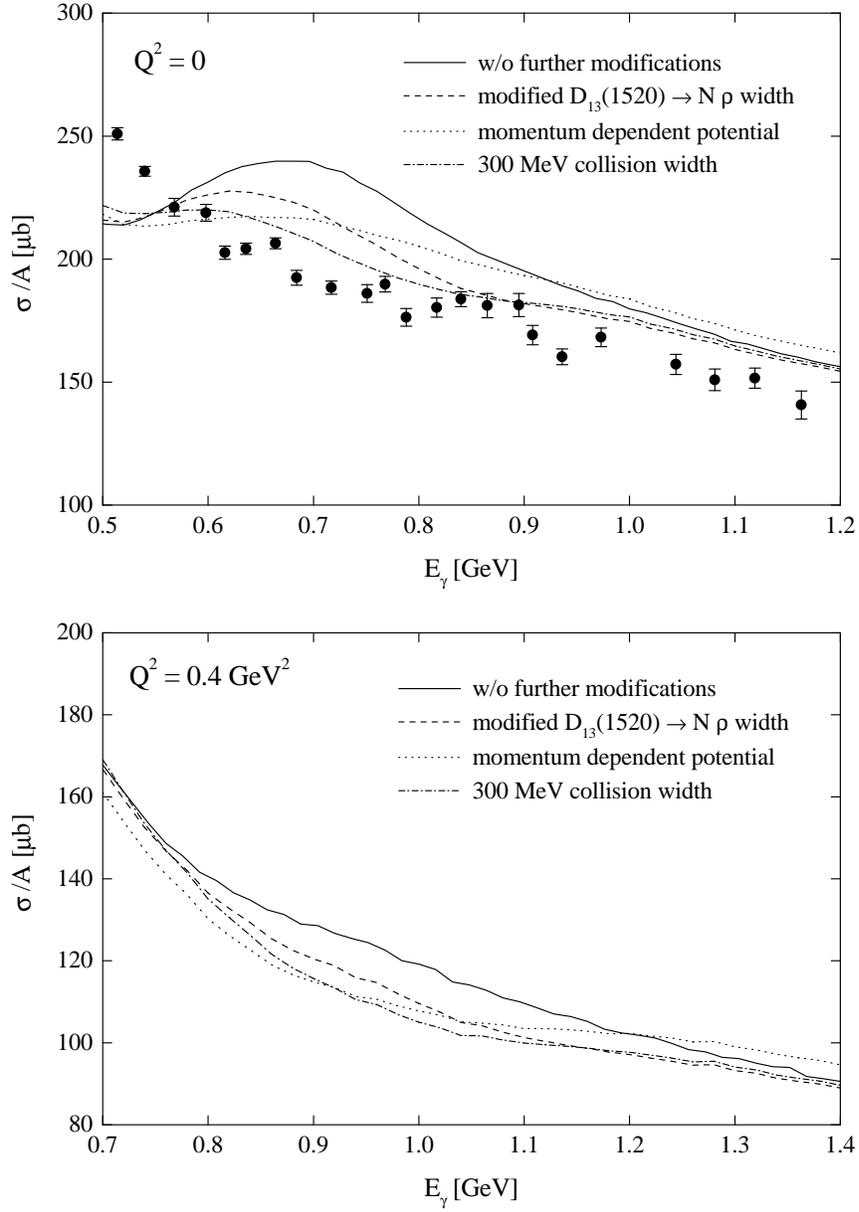,height=17cm}    
    \caption{Modifications of the second resonance region for the 
$\gamma^*{}^{40}\textrm{Ca}\to X$-cross section. 
The data are from \cite{bianchi}.}
    \label{fig:mod_1520}
  \end{center}
\end{figure}

In the following
we present different scenarios that might explain the disappearance of the 
$D_{13}(1520)$. The resonant contribution can be altered by using an enhanced 
in-medium width. This leads to a more smeared and reduced cross section.
In \cite{ef_abs} we have already proposed that the disappearance of the
$D_{13}(1520)$ might be due to its strong coupling to the $N\rho$ channel and
a medium modification of the rho meson. In \cite{post} it has
been shown within a resonance-hole model calculation for the
spectral function of the rho meson that a self-consistent treatment of the
$N\rho$ width of the $D_{13}$ indeed gives a very large broadening. An 
enhancement of the $N\rho$ width at nuclear matter density by about a factor
10 was found which gives a total width at the pole mass of about 335 MeV.
The use of this in-medium $N\rho$ width (including its full mass, momentum
and density dependence) leads to the dashed curve in fig. \ref{fig:mod_1520}.
One sees that the description of the experimental data is considerably 
improved although for photon energies around 650 MeV some bump survives. 
This bump is caused by the strong mass dependence of the
$D_{13}$ width which is also present in the in-medium width. 

The use of a momentum dependent $N^*$ potential also leads to a
smearing of the $D_{13}(1520)$, because the additional momentum dependent part
in eq. (\ref{eq:pot-mom}) leads to a shift of the effective mass.
For a $D_{13}(1520)$ produced with momenta around 800 MeV for which the 
nucleon potential almost vanishes, this shift amounts for 
$\Delta m^*\approx 50$ MeV. The 
broadening is due to the strong increase of the width with the mass. 
We observe an almost complete vanishing of the resonant structure but we still
overestimate the experimental data slightly.

An adhoc collision width $\Gamma_{coll}=300$ MeV as proposed by
\cite{kondratyuk} shows quite good agreement with the data, but is hard to 
justify.

In \cite{ochi} it has been shown that the disappearance of
the $D_{13}$ might be explained by an in-medium change of the two-pion
production cross section which results mainly from a change in the
interference structures of the different contributions due to conventional
medium modifications. However, the model in \cite{ochi} for the elementary
$\gamma N\to N\pi\pi$ process is very simple and it remains to be seen if the 
effect also shows up in an more realistic calculation.

In \cite{rapp} good agreement with the experimental data on the photoabsorption
cross section has been achieved within a model calculation of the in-medium
spectral function on the rho meson by using vector meson dominance.
We note here that the $D_{13}$ width has not been calculated 
self-consistently but the result of \cite{post} has been adopted ignoring
the mass and momentum dependence of this in-medium width, whereas we find
a substantial influence in particular of the mass dependence of the width in
our results, as shown above. Moreover, in \cite{rapp} the model parameters 
have not been adjusted to exclusive observables like $\gamma N\to \pi N$,
$\gamma N\to \pi\pi N$ as in our calculation but only to the total 
$\gamma p$ cross section. Also the neglect of an isoscalar coupling of the 
photon to the nucleon in \cite{rapp} might be questionable.

In fig. \ref{fig:mod_1520} we also show the different scenarios 
at $Q^2=0.4\ \te{GeV}^2$. Again, one can see a dropping of
the cross section in the second resonance region. The differences between the
different scenarios are rather small, because the $D_{13}(1520)$ becomes 
already very broad due to Fermi motion.

\section{Meson Production}
 
\subsection{Pion Production}

\noindent We now investigate pion production cross sections by considering
different medium modifications. For calculations involving virtual photons
we use $\varepsilon\ge0.9$.
Besides Pauli blocking and Fermi motion of the nucleons we discuss different
potentials and in-medium widths:
For the $\Delta$ resonance, we use the $N \Delta\to NN$ and 
$N\Delta\to N\Delta$ collision and Pauli blocked one-pion widths from 
\cite{ef_abs} ('standard' BUU treatment) for both the population and the final
state interactions. We discuss the effects of a momentum dependent nucleon
potential extracted from a 'medium' EOS and a momentum 
independent potential extracted from a 'hard' EOS (cf. 
sec. 2). The potentials are used for the nucleons and all resonances
except for the $\Delta$, for which we take the potential displayed in eq.
(\ref{eq:delta-pot}).

\begin{figure}[H]
  \begin{center}
\epsfig{file=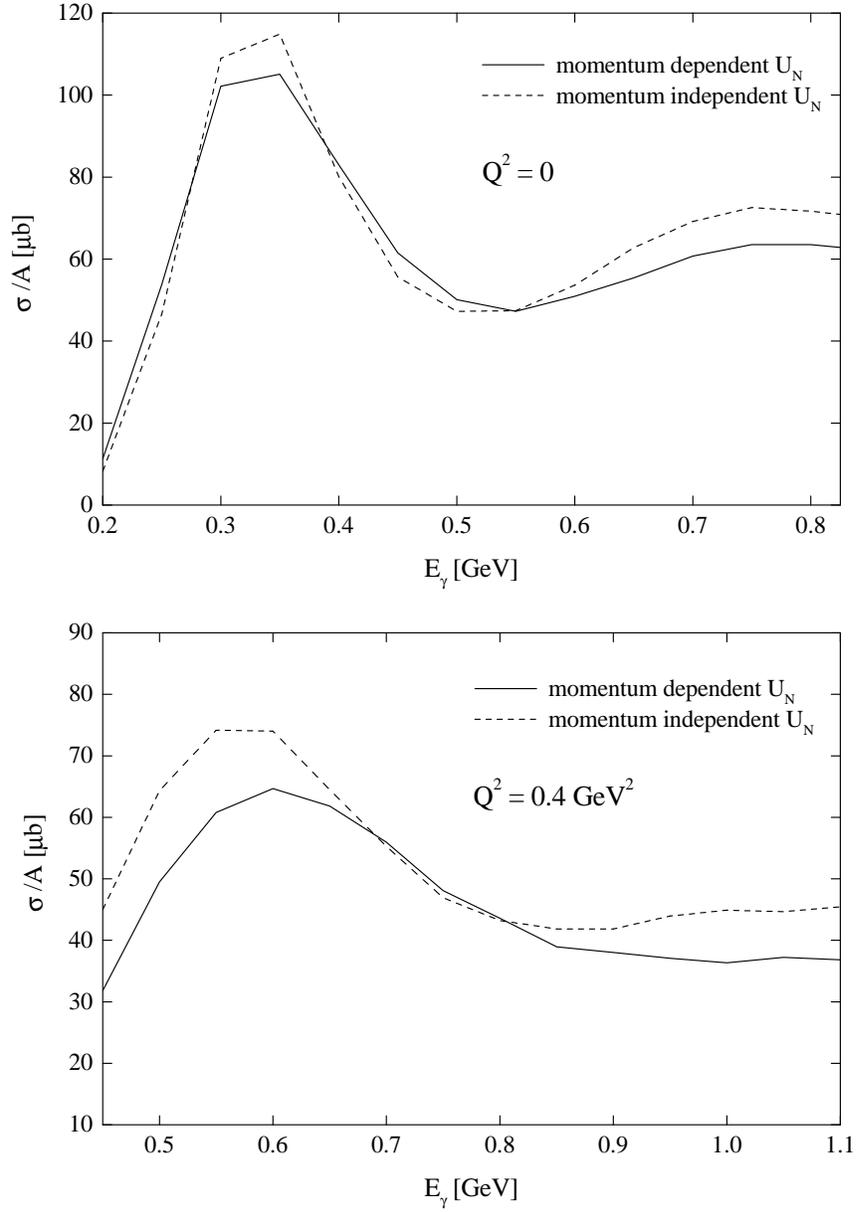,height=17cm}    
    \caption{Total cross section of the reaction 
$\gamma^*{}^{40}\textrm{Ca}\to\pi^0 X$ for $Q^2=0$ and $0.4\ \textrm{GeV}^2$.
The different curves correspond to the use of momentum-dependent and
-independent nucleon potentials.} 
    \label{fig:tsig_pi0_pot}
  \end{center}
\end{figure}

In fig. \ref{fig:tsig_pi0_pot} 
we compare both calculations for the reaction 
$\gamma^* {}^{40}\textrm{Ca}\to\pi^0 X$ for $Q^2=0$ and $0.4\ \textrm{GeV}^2$.
The momentum independent nucleon potential leads to a slight shift of the
cross section in the first resonance region to lower energies and a global 
enhancement at finite $Q^2$. This is so because at large nucleon momenta, the 
momentum 
dependent potential is larger than the momentum independent one. This leads
to the different peak maximum energies seen in fig. \ref{fig:tsig_pi0_pot}.
Furthermore, the width increases with the resonance mass, which leads to the
relative dropping of the cross section obtained with the momentum dependent
potential.

\begin{figure}[H]
  \begin{center}
\epsfig{file=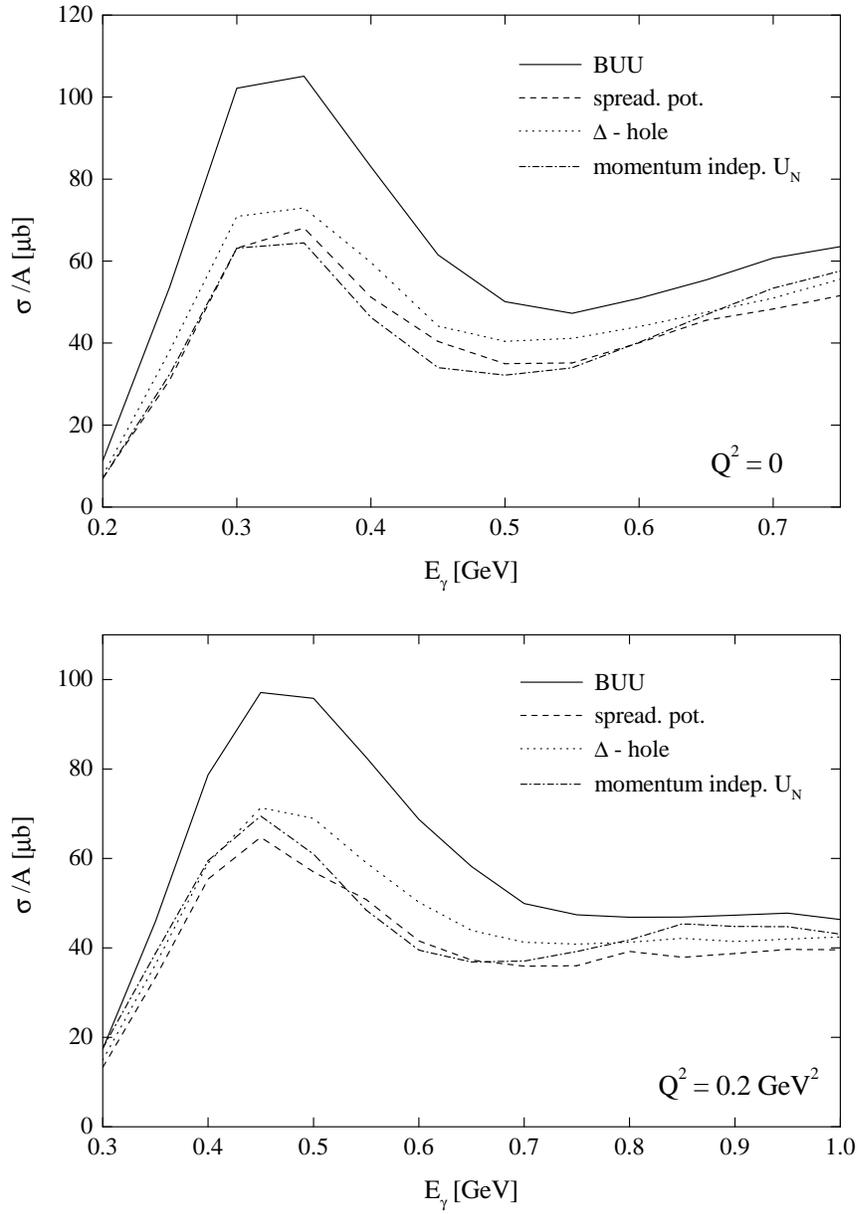,height=17cm}    
    \caption{Total cross section of the reaction 
$\gamma^*{}^{40}\textrm{Ca}\to\pi^0 X$ for different $Q^2$.}
    \label{fig:tsig_pi0_width1}
  \end{center}
\end{figure}

\begin{figure}[H]
  \begin{center}
\epsfig{file=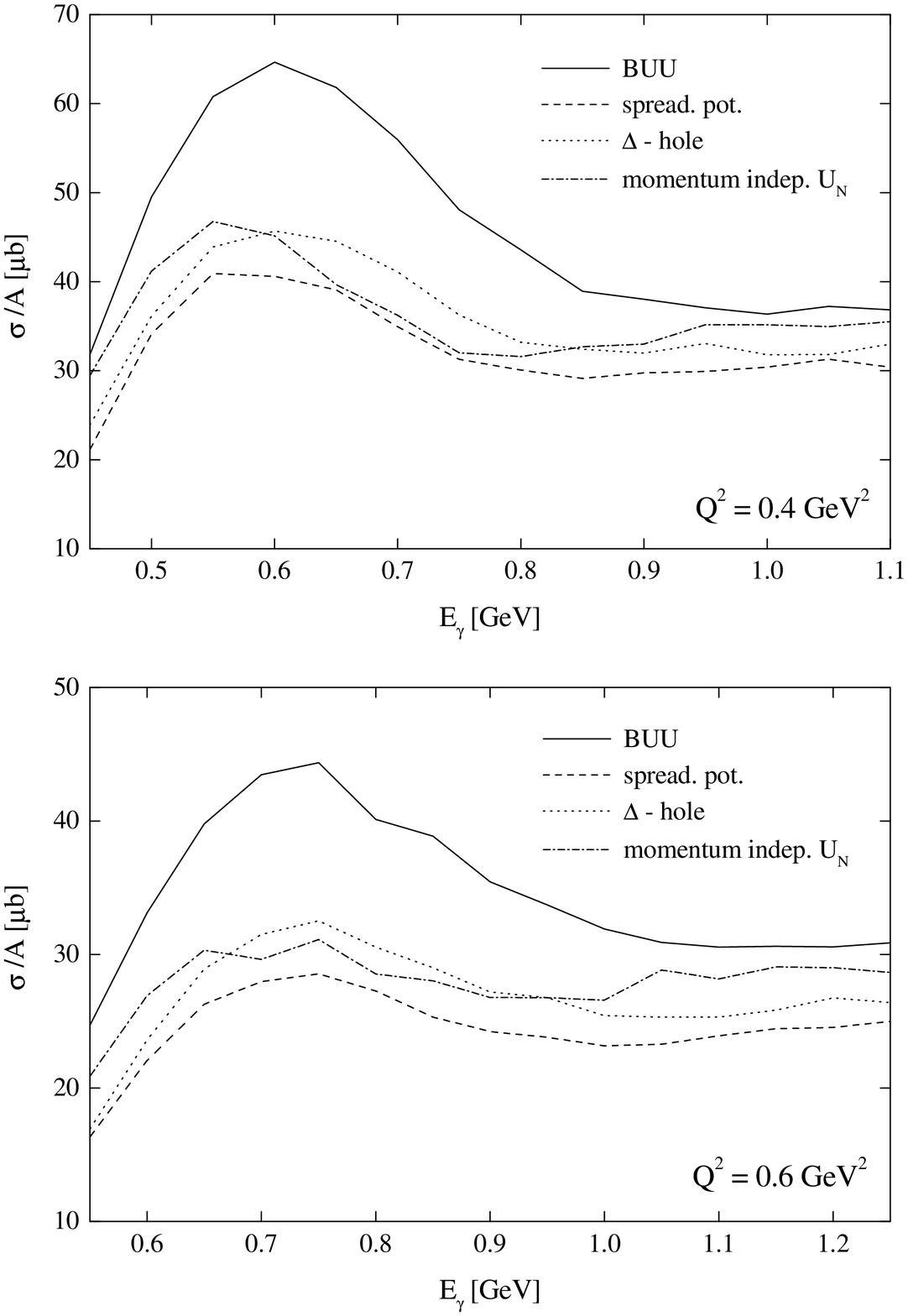,height=17cm}    
    \caption{Total cross section of the reaction 
$\gamma^*{}^{40}\textrm{Ca}\to\pi^0 X$ for different $Q^2$.
The medium modifications are explained in the text.}
    \label{fig:tsig_pi0_width2}
  \end{center}
\end{figure}

In figs. \ref{fig:tsig_pi0_width1} and \ref{fig:tsig_pi0_width2} we discuss the
different treatments of the $\Delta$ in-medium width already addressed in
sec. \ref{sec:delta-in-med} and show our results for the cross section of
the reaction $\gamma^* {}^{40}\textrm{Ca}\to\pi^0 X$ for different $Q^2$
involving the three collision widths. The respective curves are labelled by
'BUU', 'spreading potential' and '$\Delta$-hole'. All calculations have been
performed by using the momentum dependent nucleon potential mentioned 
above. In addition, we show a calculation involving the 'spreading potential'
width and the momentum independent potential.
The use of the $\Delta$-hole and spreading potential widths reduce the cross
sections by about 30\% of the BUU-width calculation, with the spreading
potential curve lying slightly below the $\Delta$-hole curve.
This can be understood, because the BUU width leads to a smaller total $\Delta$
in-medium width and thus to larger cross sections $\gamma^* N\to\Delta\to X$
(cf. eq. (\ref{eq:res_xs})). The use of the momentum independent potential
leads to the effects discussed earlier in connection with
in fig. \ref{fig:tsig_pi0_pot}.

\begin{figure}[H]
  \begin{center}
\epsfig{file=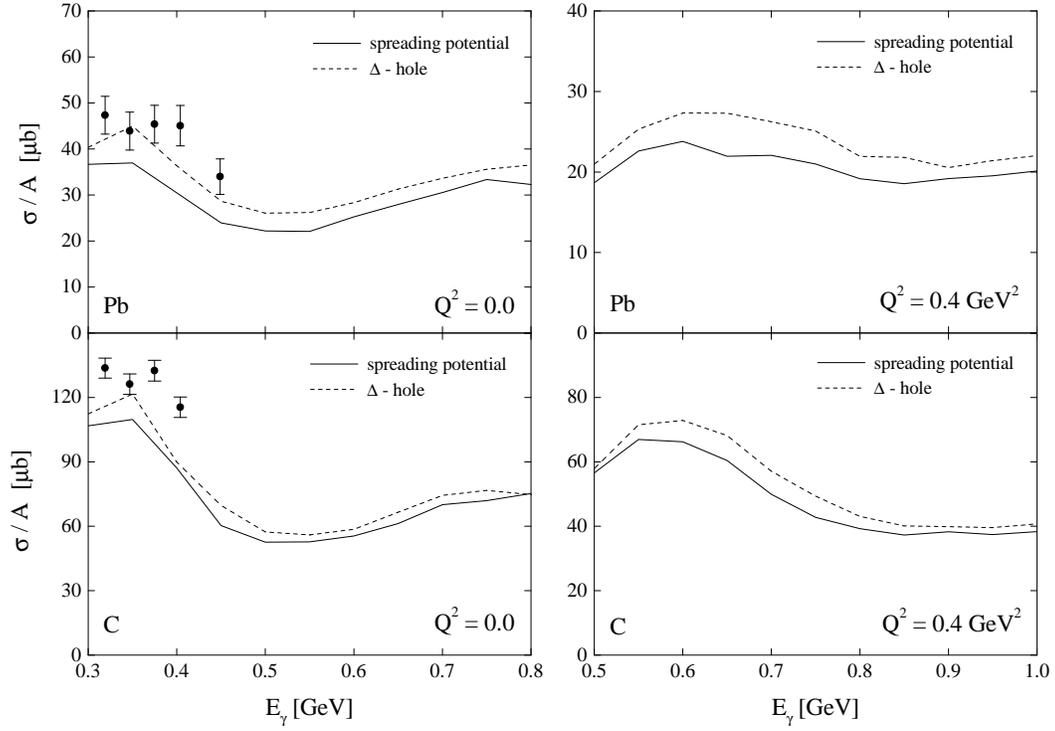,width=14cm}    
    \caption{Total cross section of the reaction 
$\gamma^* A\to\pi^0 X$ on Pb and C for $Q^2=0$ and $0.4\ \textrm{GeV}^2$.
The curves have been calculated using the spreading potential and 
$\Delta$-hole collision width. The data are from \cite{arends}.}
    \label{fig:pb_c_comp}
  \end{center}
\end{figure}

In fig. \ref{fig:pb_c_comp} we show the cross sections $\gamma^* A\to \pi^0 X$
on Pb and C, using the spreading potential and $\Delta$-hole width. For 
$Q^2=0$ we show the data from \cite{arends}. We see that the usage of these
$\Delta$ widths lead to a satisfying description of the data.

\begin{figure}[H]
  \begin{center}
\epsfig{file=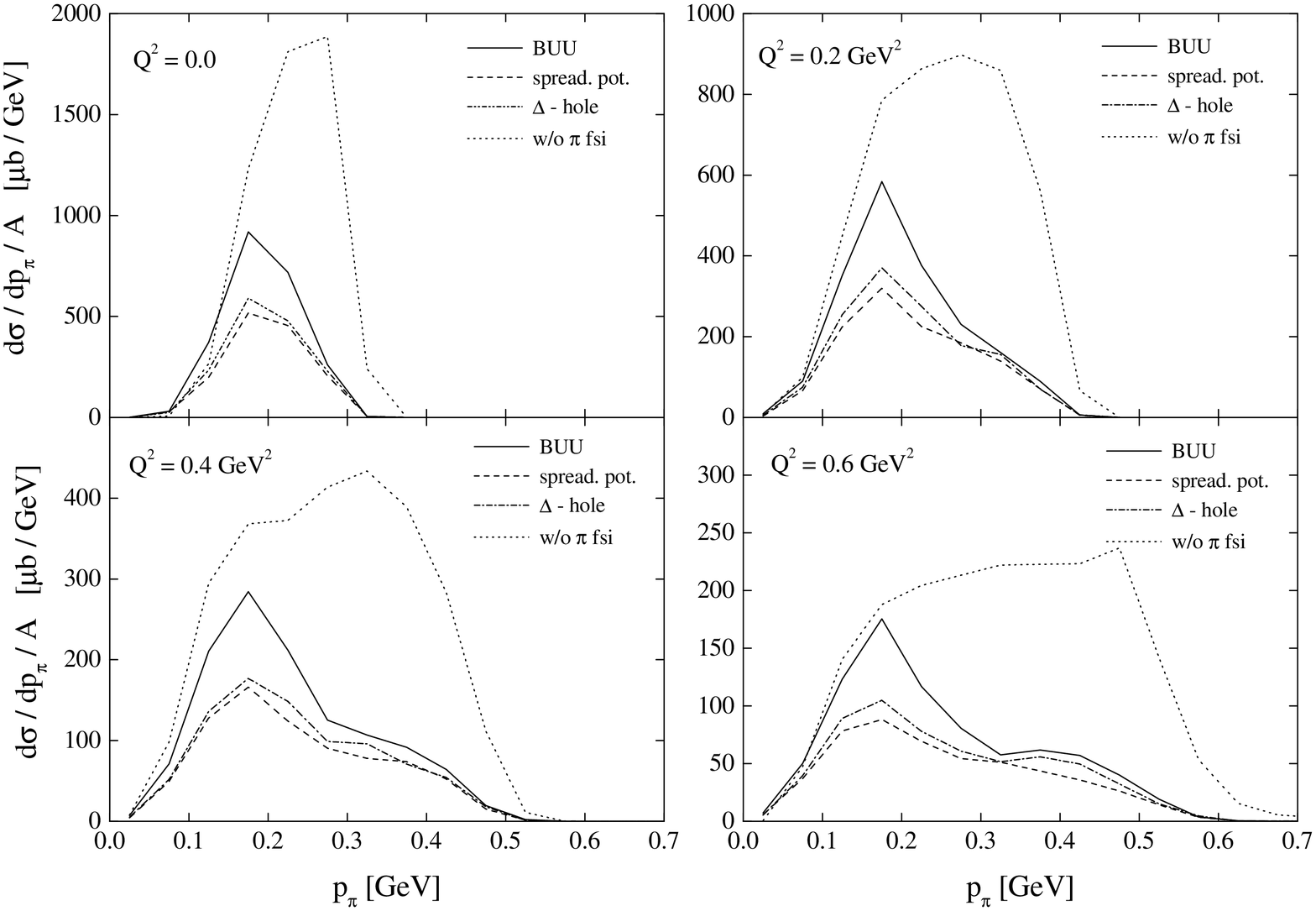,width=14cm}    
    \caption{Momentum differential cross section of the reaction
$\gamma^*{}^{40}\textrm{Ca}\to\pi^0 X$ for different $Q^2$ and 
$\sqrt s=(-Q^2+m_N^2+2 E_\gamma m_N)^{1/2}=1.23$ GeV. The medium modifications
are explained in the text.}
    \label{fig:mom_pi_325}
  \end{center}
\end{figure}

\begin{figure}[H]
  \begin{center}
\epsfig{file=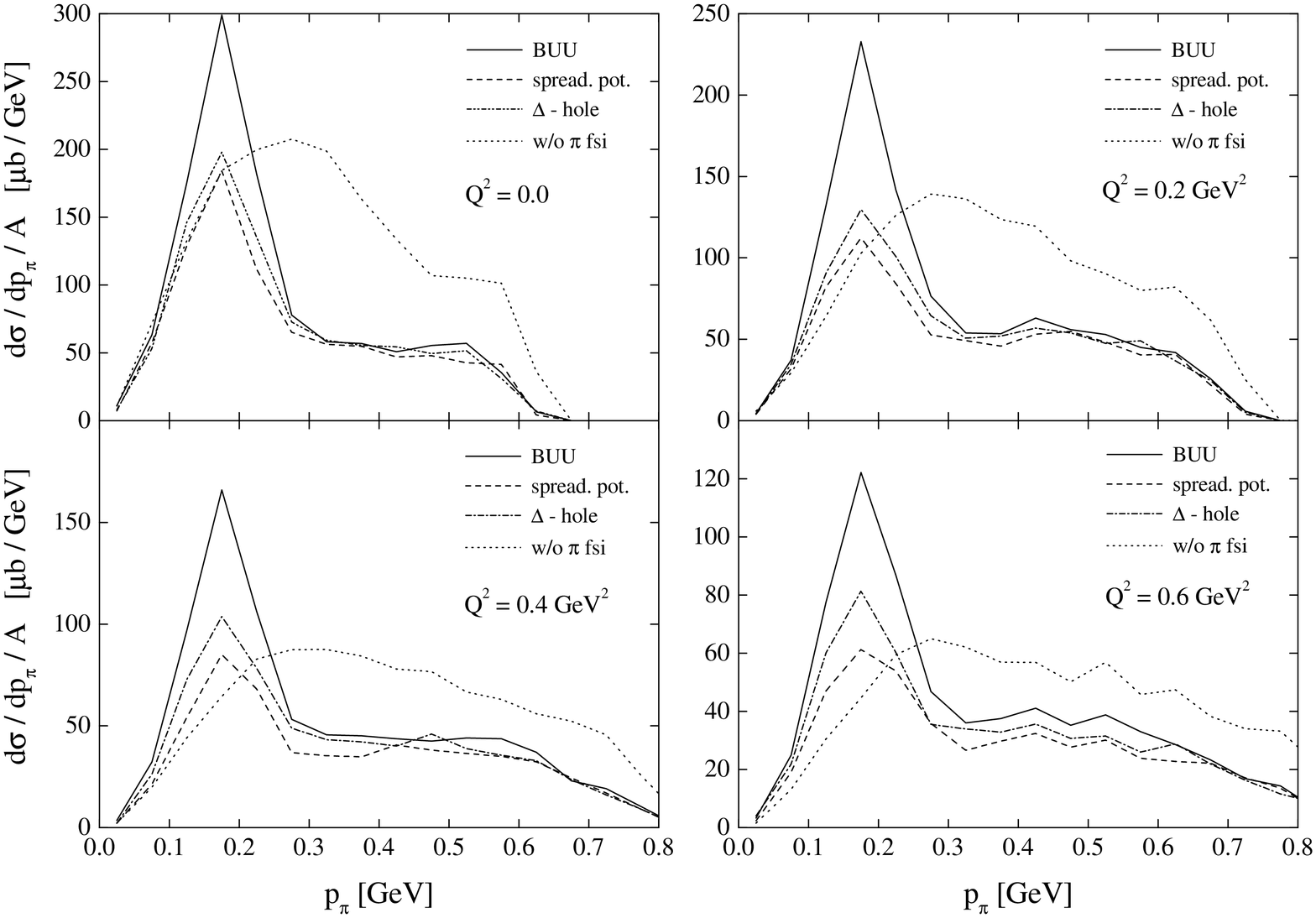,width=14cm}    
    \caption{Momentum differential cross section of the reaction
$\gamma^*{}^{40}\textrm{Ca}\to\pi^0 X$ for different $Q^2$ and
$\sqrt s=(-Q^2+m_N^2+2 E_\gamma m_N)^{1/2}=1.44$ GeV. The medium modifications
are explained in the text.}
    \label{fig:mom_pi_625}
  \end{center}
\end{figure}

In figs. \ref{fig:mom_pi_325} and \ref{fig:mom_pi_625} we show the momentum
differential cross section
$d\sigma(\gamma^* {}^{40}\te{Ca}\to\pi^0 X)/dp_\pi/A$ for different $Q^2$. 
We consider the energies $\sqrt s=1.23$ GeV, where the $\Delta$ is
excited mainly in the $\gamma^* N$ reactions and $\sqrt s=1.44$ GeV, where
the $\Delta$ excitation takes place through produced pions in the
final state interactions. Here $\sqrt s=(-Q^2+m_N^2+2E_\gamma m_N)^{1/2}$ 
denotes the cms energy involving a nucleon at rest.
We compare different calculations using the three $\Delta$ collision widths 
and also show the elementary pion
spectra obtained by suppressing all final state interactions (curve
'w/o $\pi$ fsi').
The distributions are peaked at momenta around $p_\pi=0.2$ GeV for
all $Q^2$ and both energies. The peak is due to the strong absorption of the
pions by the nucleons for momenta $p_\pi\gtrsim 300$ MeV.
For larger $Q^2$, the spectra become more smeared out compared to $Q^2=0$
due to the fact that more pions with larger momenta are produced in the 
elementary $\gamma^* N$ reactions. 
We observe significant differences between the three curves in the peak area
whereas at large momenta the differences become negligible.

The curves involving the spreading potential and $\Delta$-hole collision width
show a satisfying agreement with the data for $Q^2=0$ from the TAPS 
collaboration \cite{krusche_privat,krusche_pol}, which holds for both the 
shape and the magnitude of the spectra.

\subsubsection{Effects of Medium Modifications of the $D_{13}(1520)$ 
Resonance}

\begin{figure}[H]
  \begin{center}
\epsfig{file=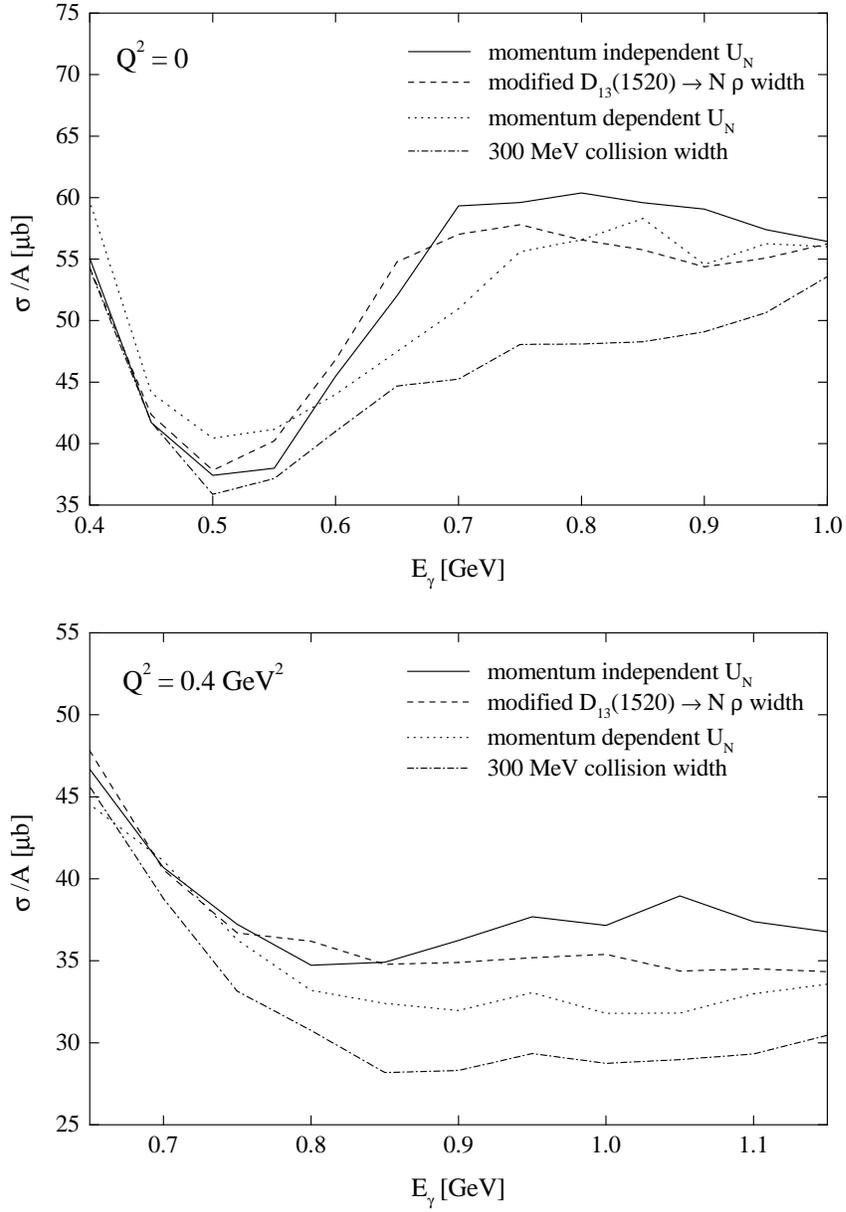,height=17cm}    
    \caption{Modifications to the cross section of the reaction
$\gamma^*{}^{40}\textrm{Ca}\to\pi^0 X$ in the second resonance region for
$Q^2=0$ and $0.4\ \textrm{GeV}^2$. For explanations see text.}
    \label{fig:tsigma_1520}
  \end{center}
\end{figure}

In fig. \ref{fig:tsigma_1520} we discuss the influence of the medium 
modifications of the $D_{13}(1520)$ already considered in the last section
on the reaction $\gamma^* {}^{40}\te{Ca}\to\pi^0 X$
for $Q^2=0$ and $Q^2=0.4\ \te{GeV}^2$. The curve labelled with
'momentum independent $U_N$' refers to a calculation using the $\Delta$-hole
collision width and the $\Delta$ potential from eq.
(\ref{eq:delta-pot}) with the momentum independent nucleon potential.
The additional use of a 300 MeV collision
width for both population and decay of the $D_{13}(1520)$ clearly leads to a 
strong reduction of the cross section in the second resonance region. 
Using the modified $N\rho$ decay width described in the last section instead,
we observe a reduction of the cross section. Now rho mesons are 
produced at a higher rate, but most of them contribute to pion production via 
direct decay or excitation of an intermediate resonance with subsequent pion 
decay.
A calculation using the momentum dependent nucleon potential instead
of $D_{13}(1520)$ modifications also results in a dropping of the cross 
section. In the second resonance region the additional momentum dependent
term leads already for $Q^2=0$ to a significantly larger effective resonance
mass, which influences the cross section through the changed total width.

\begin{figure}[H]
  \begin{center}
\epsfig{file=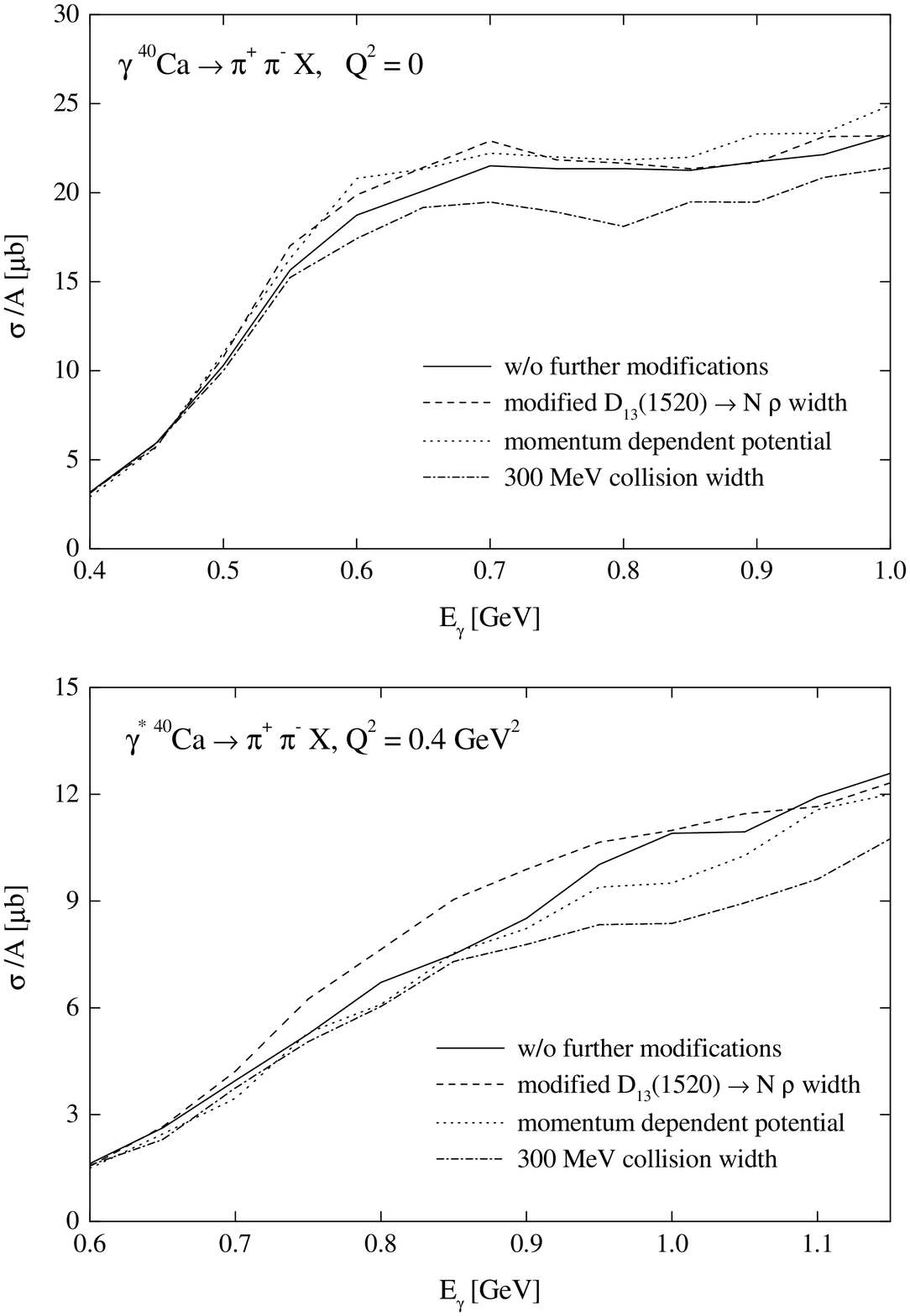,height=17cm}    
    \caption{Cross section of the reaction 
$\gamma^*{}^{40}\textrm{Ca}\to\pi^+ \pi^- X$ for different $Q^2$. The
medium modifications are explained in the text.}
    \label{fig:pip_pim_1520}
  \end{center}
\end{figure}

\begin{figure}[H]
  \begin{center}
\epsfig{file=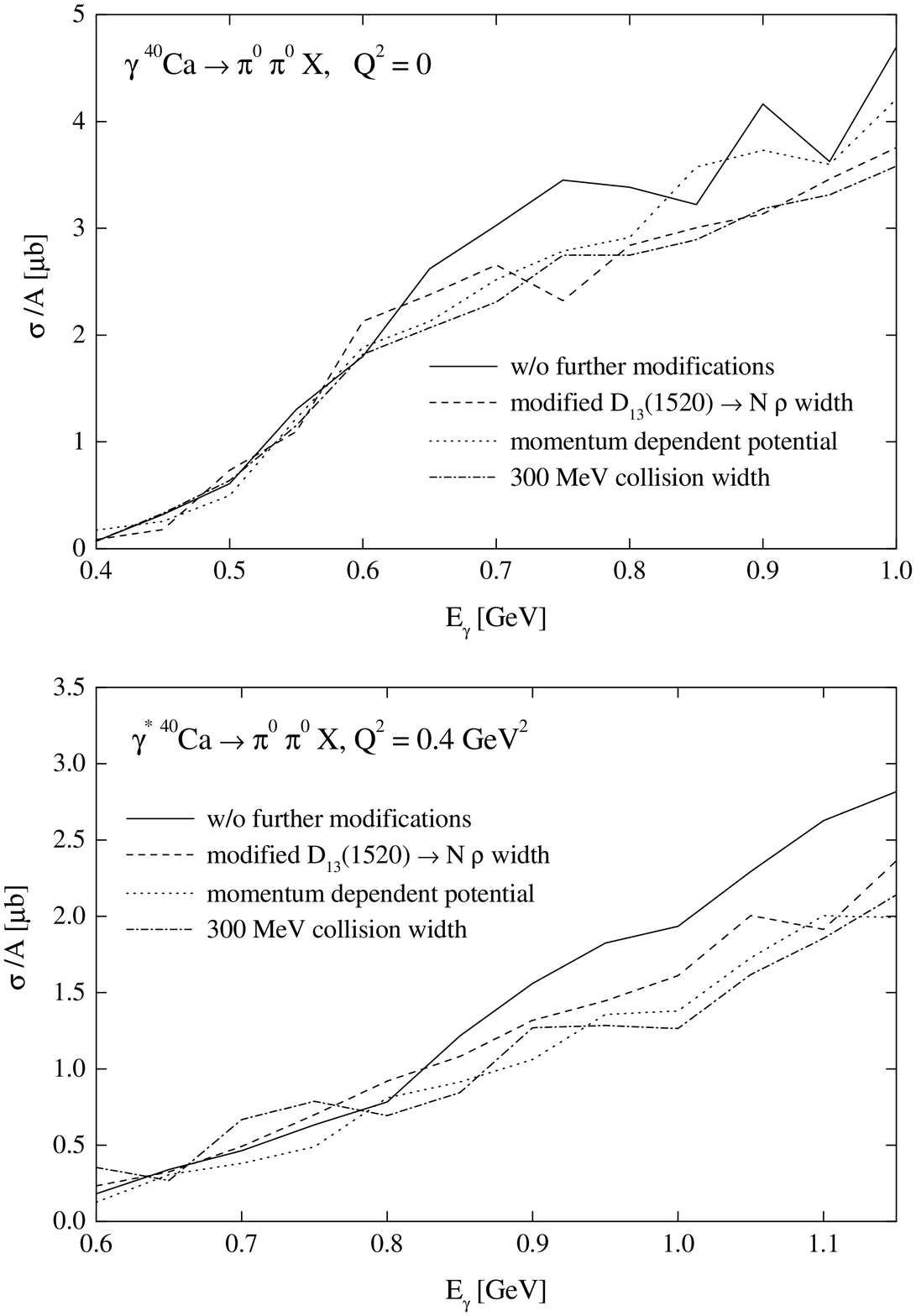,height=17cm}    
    \caption{Cross section of the reaction 
$\gamma^*{}^{40}\textrm{Ca}\to\pi^0 \pi^0 X$ for different $Q^2$. The
medium modifications are explained in the text.}
    \label{fig:pi0_pi0_1520}
  \end{center}
\end{figure}

In figs. \ref{fig:pip_pim_1520} and \ref{fig:pi0_pi0_1520} we show our
results for the cross sections $\gamma^* {}^{40}\te{Ca}\to\pi^+ \pi^- X$
and $\gamma^* {}^{40}\te{Ca}\to\pi^0 \pi^0 X$. For $Q^2=0$ we see a strong
increase of the cross sections at about 0.5 GeV, whereas at energies above
0.7 GeV it varies only weakly. For $Q^2=0.4\ \textrm{GeV}^2$ this behaviour
is washed out. The different medium modifications of the $D_{13}(1520)$
lead to very similar results.

\subsection{Eta Production}

\begin{figure}[H]
  \begin{center}
\epsfig{file=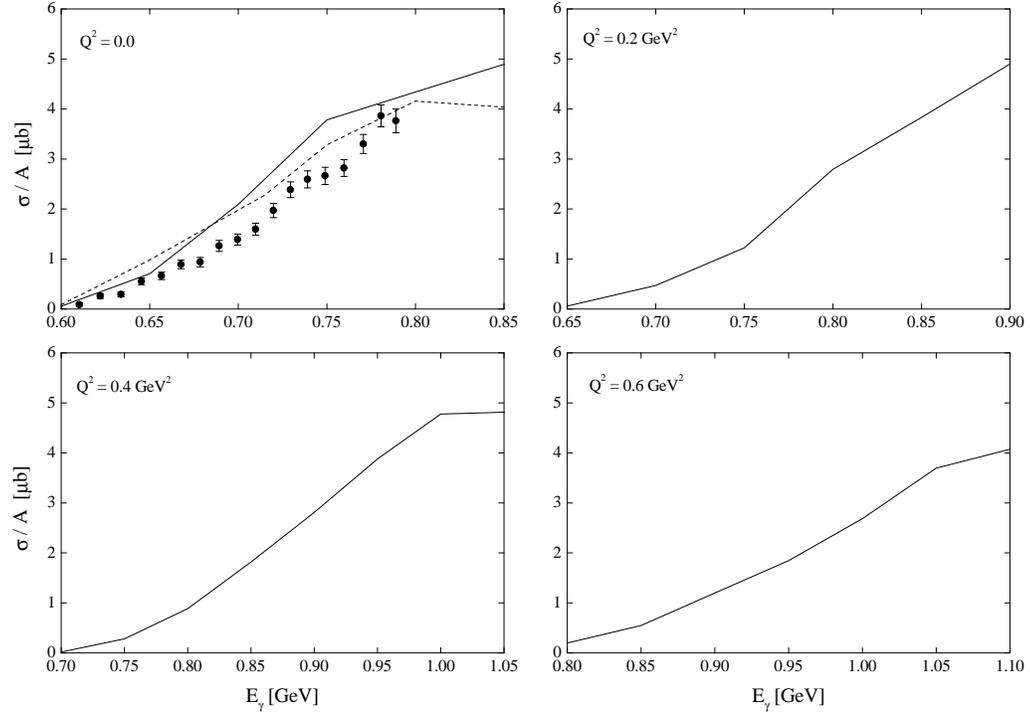,width=14cm}    
    \caption{Cross section of the reaction 
$\gamma^*{}^{40}\textrm{Ca}\to\eta X$ 
for different $Q^2$. For $Q^2=0$ the dashed curve was calculated using the
constant $\eta N$ cross sections.
The data for
$Q^2=0$ are from \cite{rlandau}.}
    \label{fig:eta_ca40_tot}
  \end{center}
\end{figure}

\noindent  The investigation of eta 
production cross sections gives insight into the $\eta N$ dynamics and the
behaviour of the $S_{11}(1535)$ in nuclear matter, since all primarily 
produced etas stem from the excitation of a $S_{11}(1535)$ resonance.
As we have shown in \cite{ef_prod}, the cross sections are dominated
by the primary etas, the contributions from secondary processes such as
$\pi N\to\eta N$ are negligible.
In fig. \ref{fig:eta_ca40_tot} we show the total cross section of the
reaction $\gamma^* {}^{40}\textrm{Ca}\to\eta X$, where
one observes a broadening and dropping of the cross section with increasing
$Q^2$. For $Q^2=0$ we obtain good agreement with the data from
\cite{rlandau} (see also \cite{ef_prod}, where the agreement was slightly
better; the difference between the two calculations is due to systematic
uncertainties in the transport calculation, such as initialization and
resonance properties).

\begin{figure}[H]
  \begin{center}
\epsfig{file=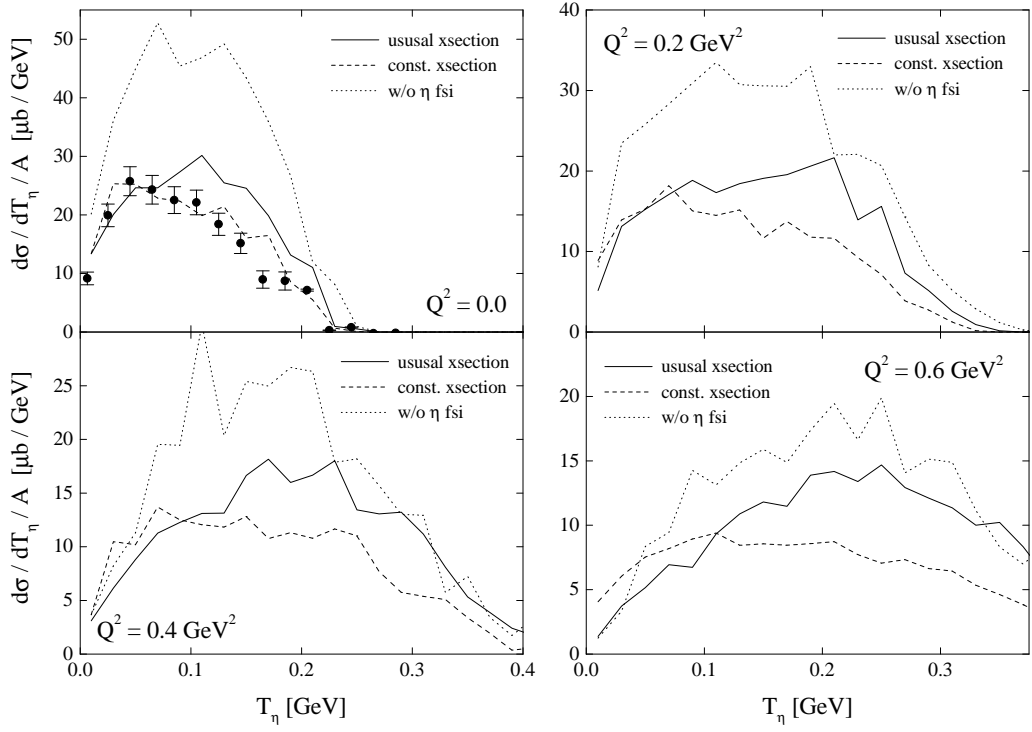,width=14cm}    
    \caption{Energy differential cross section of the reaction
$\gamma^*{}^{40}\textrm{Ca}\to\eta X$ for different $Q^2$ for $\sqrt s=1.54$
GeV. The different curves are explained in the text. The data are from
\cite{rlandau}.}
    \label{fig:eta_tksig}
  \end{center}
\end{figure}

In fig. \ref{fig:eta_tksig} we show our results for the energy differential
cross section $d\sigma(\gamma^*{}^{40}\te{Ca}\to\eta X)/dT_\eta/A$ for 
$\sqrt s=1.54$ GeV and different $Q^2$. The data from \cite{rlandau} for 
$Q^2=0$ are also shown. The calculation labelled by
'usual xsection' involved the momentum independent nucleon potential.
The shift of the
calculated spectrum to higher eta energies at $Q^2=0$ compared to the data
was discussed in 
\cite{ef_sibi}. There it was shown, that the structure of the spectra could
be explained by using the energy independent cross sections 
$\sigma(\eta N\to\eta N)=20$ mb and $\sigma(\eta N\to\pi N)=30$ mb for
the final state interactions; for higher eta kinetic energies
these values are larger than those one obtains from an interaction through
the $S_{11}(1535)$ alone.
The corresponding curves (labelled by 'constant xsection') are
also shown. The spectra are reduced at larger eta kinetic energies,
because in this range the constant inelastic cross section exceeds the 
cross section used in the usual resonance model, which leads to a relative
loss of etas. Also shown are the eta spectra obtained by suppressing all final 
state interactions (curve 'w/o $\eta$ fsi').

For increasing $Q^2$ we observe that the maximum of the spectrum moves
towards higher kinetic energies. A investigation at different $\sqrt s$ leads 
to the same result. In the pion spectra in figs. \ref{fig:mom_pi_325} and
\ref{fig:mom_pi_625} such a shift of the 
distributions was not observed. This behaviour can be explained, if one recalls
that the detected eta mesons do not undergo final state interactions at such a
degree as the pions. Therefore, the eta spectrum is mainly due to the emission
spectrum of the $S_{11}(1535)$. This is supported by the fact that the curves
calculated without final state interactions show the same 
structure as the physical eta spectra. For large $Q^2$ the momentum transfer
increases and so does the average momentum of the initially produced
$S_{11}(1535)$ and its decay products, which shifts the eta spectrum.
The measurement of eta electroproduction may therefore clarify whether the
$\eta N-S_{11}(1535)$ dynamics is treated correctly in our model.

\section{Summary}

\noindent We have presented calculations of inclusive electroproduction and
photoproduction processes on nuclei in the resonance region within a model
that can take coupled channel effects in the final state interactions into
account.

The cross section for the $\gamma^* N$ processes have been parametrized by
using the $\gamma N$ cross sections of the respective channels to describe
the energy dependence and introducing form factors for the 
$Q^2$-dependence. These form factors have been obtained by 
fits to data in different channels.

In addition to \cite{ef_abs}, we have discussed modifications of the 
photoabsorption cross section on nuclei for the $D_{13}(1520)$,
which led to a better description of the data in the second resonance region.
The results for $\gamma^* A\to X$ processes at finite $Q^2$ are found to be
strongly 
dominated by the Fermi motion of the nucleons which leads to a 
disappearance of the second resonance region for $Q^2\ge 0.2\ \te{GeV}^2$.
Other medium modifications like Pauli blocking, in-medium widths and 
$\Delta$ potential do not show much influence at large $Q^2$.
The comparison with data in the $\Delta$ region above the $\Delta$ peak has
shown 
very good agreement for several kinematics of the scattered electrons.

Turning to meson production on nuclei, we first discussed several treatments
of the $\Delta$ width, originating from the phenomenological spreading 
potential
and calculations of the $\Delta$ self energy, which included the absorption
of $\Delta$s on more than one nucleon. The comparison with the approach in
\cite{ef_prod} has shown smaller total and momentum-differential cross 
sections.
The discussion of the momentum-differential $\pi^0$-cross section showed
a peak at pion momenta around 0.2 GeV, remaining there for increasing $Q^2$.
The energy differential cross section for eta production has shown a  
distribution moving towards higher energies for increasing $Q^2$. 
\medskip

Additional measurements of meson electroproduction on nuclei are inevitable, 
because this
could answer different questions, e.g. whether the $Q^2$-dependences of the
elementary
gamma-nucleon processes are chosen correctly; also aspects concerning
the resonance and meson dynamics in nuclei could be clarified.

\newpage

\newpage

\newpage

\newpage

\newpage

\newpage

\newpage

\newpage

\newpage

\newpage


\newpage

\newpage

\newpage

\newpage


\newpage

\newpage

\newpage

\newpage

\newpage


\newpage

\newpage

\begin{table}[H]
  \begin{center}
\setlength\extrarowheight{4pt}
    \begin{tabular}{|c|c|c|c|}
     \hline
       & ${}^{12}\textrm{C}$ & ${}^{40}\textrm{Ca}$ & ${}^{208}\textrm{Pb}$
\\\hline 
     $r_0$ [fm] & 2.209 & 3.614 & 6.755 \\\hline
     $\alpha$ [fm] & 0.479 & 0.479 & 0.476\\\hline     
    \end{tabular}
    \caption{Parameters $r_0,\alpha$ for the Woods-Saxon distribution.}
    \label{tab:parameters}
  \end{center}
\end{table}

\begin{table}[H]
  \begin{center}
\setlength\extrarowheight{4pt}
    \begin{tabular}{|c|c||c|c|c|c|c|c|}
\hline
 & $\varepsilon$ bin & $P_{33}(1232)$ & $D_{13}(1520)$ & $S_{11}(1535)$ & $F_{15}(1680)$
 & $1\pi$ & $2\pi$\\
\hline\hline
   & $\varepsilon\ge0.9$   & 1.06 &      &      & 0.7 &     & 0.6\\\cline{2-3}
\cline{6-6}
$a$& $0.6<\varepsilon<0.9$ & 1.0  & 0.65 & 1.07 & 0.6 & 1.0 &    \\\cline{2-3}
\cline{6-6}\cline{8-8}
   & $\varepsilon\le0.6$   & 1.03 &      &      & 0.45&     & 0.65\\
\hline\hline
   & $\varepsilon\ge0.9$   & 1.85 &      &      &     &     &     \\\cline{2-2}
$b$& $0.6<\varepsilon<0.9$ &      & 1.25 & 3.0  & 2.0 & 2.5 & 1.95\\\cline{2-3}
   & $\varepsilon\le0.6$   & 1.8  &      &      &     &     &     \\
\hline\hline
   & $\varepsilon\ge0.9$   &      &      &      &     &     & \\\cline{2-2}
$c$& $0.6<\varepsilon<0.9$ & 1.0  & 0.5  & 0.5  & 0.5 & 0.5 & 0.975\\\cline{2-2}
   & $\varepsilon\le0.6$   &      &      &      &     &     & \\
\hline
    \end{tabular}
    \caption{Parameters $a,b$ und $c$ for the form factors.}
    \label{tab:ff_par}
  \end{center}
\end{table}


\begin{thebibliography}{1}
\bibitem{koch}{C.M. Ko, V. Koch and G.Q. Li, Ann. Rev. Nucl. Part. Sci. 47
(1997), 505}
\bibitem{agakichiev}{G. Agakichiev et al., Phys. Rev. Lett. 75 (1995), 1272} 
\bibitem{mazzoni}{M.A. Mazzoni, Nucl. Phys. A 566 (1994), 95c; M. Masera, Nucl.
 Phys. A 590 (1995), 93c}
\bibitem{bianchi}{N. Bianchi et al., Phys. Rev. C54 (1996), 1688}
\bibitem{post}{W. Peters, M. Post, H. Lenske, S. Leupold and
U. Mosel, Nucl. Phys. A632 (1998), 109}
\bibitem{barreau}{P. Barreau et al., Nucl. Phys. A402 (1983), 515} 
\bibitem{sealock}{R.M. Sealock et al., Phys. Rev. Lett. 62 (1989), 1350} 
\bibitem{gil}{A. Gil, J. Nieves and E. Oset, Nucl. Phys. A627 (1997), 543} 
\bibitem{gil2}{A. Gil, J. Nieves and E. Oset, Nucl. Phys. A627 (1997), 599} 
\bibitem{ef_abs}{M. Effenberger, A. Hombach, S. Teis and U. Mosel, 
Nucl. Phys. A613 (1997), 353}
\bibitem{ef_prod}{M. Effenberger, A. Hombach, S. Teis and U. Mosel, 
Nucl. Phys. A614 (1997), 501}
\bibitem{ef_dil}{M. Effenberger, E.L. Bratkovskaya and U. Mosel, 
nucl-th/9903026}
\bibitem{teis}{S. Teis et al., Z. Phys. A356 (1997), 421}
\bibitem{hombach}{A. Hombach, W. Cassing, S. Teis and U. Mosel, 
nucl-th/9812050, Eur. Phys. J. A, in press} 
\bibitem{ef_pi}{M. Effenberger, E.L. Bratkovskaya, W. Cassing and U. Mosel, 
nucl-th/9901039}
\bibitem{anderson}{B. Anderson, G. Gustafson and Hong Pi, Z. Phys. C57 (1993),
485}
\bibitem{cassing}{W. Cassing, V. Metag, U. Mosel and K. Niita, 
Phys. Rep. 188 (1990), 363} 
\bibitem{hirata}{M. Hirata, J.H. Koch, F. Lenz and E.J. Moniz, 
Ann. Phys. 120 (1979), 205}
\bibitem{weise}{T. Ericson and W. Weise, Pions and Nuclei, Clarendon Press,
 Oxford 1988}
\bibitem{oset_delta}{E. Oset and L.L Salcedo, Nucl. Phys. A468 (1987), 631}
\bibitem{stoler}{P. Stoler, Phys. Rep. 226 (1993), 104}
\bibitem{brasse}{F.W. Brasse et al., Nucl. Phys. B110 (1976), 413}
\bibitem{arndt}{R.A. Arndt, R.L. Workman, Z. Li and L.D. Roper, 
Phys. Rev. C42 (1990), 1853}
\bibitem{manley}{D.M. Manley and E.M. Saleski, Phys. Rev. D45 (1992), 4002}
\bibitem{walker}{R.L. Walker, Phys. Rev. 182 (1969), 1729}
\bibitem{warns}{M. Warns, H. Schr$\ddot\te{o}$der, W. Pfeil and H. Rollnik,
Z. Phys. C45 (1990), 627}
\bibitem{carrasco}{R.C. Carrasco and E. Oset, Nucl. Phys. A536 (1992), 445}
\bibitem{ochi}{M. Hirata, K. Ochi and T. Takaki, Phys.Rev.Lett. 80 (1998), 
5068}
\bibitem{kondratyuk}{L.A. Kondratyuk, M.I. Krivoruchenko, N. Bianchi,
E. De Sanctis and V. Muccifora, Nucl. Phys. A579 (1994), 453}
\bibitem{rapp}{R. Rapp, M. Urban, M. Buballa and J. Wambach, 
Phys. Lett. B417 (1998), 1; R. Rapp, nucl-th/9804065}

\bibitem{arends}{J. Arends et al., Nucl. Phys. A454 (1986), 579}
\bibitem{krusche_privat}{B. Krusche, private communication and to be published}
\bibitem{krusche_pol}{B. Krusche, Acta Phys. Pol. B29 (1998), 3335}
\bibitem{rlandau}{M.E. R$\ddot\te{o}$big-Landau et al., Phys. Lett.B373 (1996),
45} 
\bibitem{ef_sibi}{M. Effenberger and A. Sibirtsev, 
Nucl. Phys. A632 (1998), 99} 


 
\bibitem{sh72}{W.J. Shuttleworth et al., Nucl. Phys. B45 (1972), 428}
\bibitem{si71}{R. Siddle et al., Nucl. Phys. 35 (1971), 93}
\bibitem{al76}{J.C. Alder et al., Nucl. Phys. B105 (1976), 253}
\bibitem{la81}{A. Latham et al., Nucl. Phys. B189 (1981), 1}
\bibitem{ga72}{S. Galster et al., Phys. Rev. D5 (1972), 519}
\bibitem{brasse78}{F.W. Brasse et al., Nucl. Phys. B139 (1978), 37}
\bibitem{armstrong}{T.A. Armstrong et al., Phys. Rev. D5 (1972), 1640}
\bibitem{baldini}{Baldini, Flamino, Moorhead and Morrison, 
Landolt-B$\ddot\te{o}$rnstein, Band 12, Springer Verlag, Berlin 1987}
\bibitem{meziani}{Z.E. Meziani et al., Phys. Rev. Lett. 54 (1985), 1233}
\bibitem{zgiche}{A. Zghiche et al., Nucl. Phys. A572 (1994), 513}
\bibitem{amaldi}{E. Amaldi, Pion-Electroproduction, Springer Tracts in
modern physics 83, Berlin 1979}
\bibitem{stein}{S. Stein et al., Phys. Rev. D12 (1975), 1884}
\bibitem{krusche}{B. Krusche et al., Phys. Rev. Lett. 74 (1995), 3736}


\end{thebibliography}
\end{document}